\documentclass[sigconf]{acmart}
\setcopyright{none}
\renewcommand\footnotetextcopyrightpermission[1]{}
\usepackage{amsmath,amssymb,amsfonts}
\usepackage{amsthm}
\usepackage{circledsteps}
\usepackage{array}
\usepackage{algorithm}
\usepackage{algorithmicx}
\usepackage{float}

\usepackage{algpseudocode}
\usepackage{graphicx}
\usepackage{caption}
\usepackage{subcaption}
\usepackage{tabularx}
\usepackage{textcomp}
\usepackage{xcolor}
\def\BibTeX{{\rm B\kern-.05em{\sc i\kern-.025em b}\kern-.08em
    T\kern-.1667em\lower.7ex\hbox{E}\kern-.125emX}}
    
\usepackage{etoolbox}
\usepackage{flushend}
\usepackage{bm}
\usepackage{appendix}

\hypersetup{breaklinks=true}

\usepackage{hyperref}
\usepackage{comment}

\newcommand{\sysname}{CoMesh}

\newcommand{\kgroup}{$k$-group}
\newcommand{\kgroups}{$k$-groups}
\newcommand{\smart}{smart device}

\newcommand{\K}{k}
\newcommand{\f}{f}
\newcommand{\F}{F}

\newtheorem{definition}{Definition}
\newtheorem{lemma}{Lemma}
\newtheorem{theorem}{Theorem}

\newcommand{\Figure}{Fig.}

\newcommand{\Section}{Sec.}

\newcommand{\Safety}{Safety}
\newcommand{\Liveness}{Liveness}

\newcommand{\Table}{Table}

\newtheorem{innercustomgeneric}{\customgenericname}
\providecommand{\customgenericname}{}
\newcommand{\newcustomtheorem}[2]{%
  \newenvironment{#1}[1]
  {%
   \renewcommand\customgenericname{#2}%
   \renewcommand\theinnercustomgeneric{##1}%
   \innercustomgeneric
  }
  {\endinnercustomgeneric}
}

\newcustomtheorem{customdef}{Definition}
\newcustomtheorem{customlemma}{Lemma}
\newcustomtheorem{customthm}{Theorem}

\algnewcommand\And{\textbf{and}\space}
\algnewcommand\Or{\textbf{or}\space}
\algnewcommand\Not{\textbf{not}\space}
\algnewcommand\True{\textbf{true}\space}
\algnewcommand\False{\textbf{false}\space}
\algnewcommand\algorithmicforeach{\textbf{for each}}
\algdef{S}[FOR]{ForEach}[1]{\algorithmicforeach\ #1\ \algorithmicdo}

\newcommand{\squishlist}
{
    \begin{list}{$\bullet$}
    {
        \setlength{\itemsep}{0pt}      \setlength{\parsep}{3pt}
        \setlength{\topsep}{3pt}       \setlength{\partopsep}{0pt}
        \setlength{\leftmargin}{1.5em} \setlength{\labelwidth}{1em}
        \setlength{\labelsep}{0.5em}
    }
}
\newcommand{\squishend}
{
    \end{list}
}

\begin{document}

\title{\sysname{}: Fully-Decentralized Control for Sense-Trigger-Actuate Routines in Edge Meshes}

\thanks{Work supported in part by  NSF CNS 1908888, and a Capital One gift.}

\author{Anna Karanika, Rui Yang, Xiaojuan Ma, Jiangran Wang, Shalni Sundram, Indranil Gupta}
\email{{annak8,ry2,xm20,jw22,shalnis2,indy}@illinois.edu}
\affiliation{%
    \department{Department of Computer Science}
    \institution{University of Illinois at Urbana-Champaign}
    \city{Urbana}
    \state{IL}
    \country{USA}}

\begin{abstract}

While mesh networking for edge settings (e.g., smart buildings, farms, battlefields, etc.) has received much attention, the layer of control over such meshes remains largely centralized and cloud-based. This paper focuses on applications with sense-trigger-actuate (STA) workloads---these are similar to the abstraction of routines popular in smart homes, but applied to larger-scale edge IoT deployments. We present \sysname, which tackles the challenge of building local, non-cloud, and decentralized solutions for control of sense-trigger-actuate  applications. At its core \sysname{} uses an abstraction called k-groups to spread in a fine-grained way, the load of STA actions. Coordination within the k-group uses selective fast and cheap mechanisms rather than expensive off-the-shelf solutions. k-group selection is proactively dynamic, and occurs by using a combination of zero-message-exchange mechanisms (to reduce load) and locality sensitive hashing (to be aware of physical layout of devices). We analyze and theoretically prove the safety of \sysname's mechanisms. Our evaluations using both simulation and Raspberry Pi lab deployments show that \sysname{} is load-balanced, fast, and fault-tolerant.

\end{abstract}



\maketitle
% Temporarily add this to show page number.
\thispagestyle{plain}
\pagestyle{plain}

\section{Introduction}
\label{section:intro}

Commercial edge deployments with Internet of Things (IoT) devices are rapidly growing in size and density. There are over 10 B IoT devices today, expected to exceed 25 B by 2030~\cite{IoTstats}. For instance, smart buildings account for 1.26 B devices, and the smart building IoT sector is forecast to grow to 2.5 B devices by 2027, and comprise a \$90+ B market~\cite{Memoori}.
Other emerging areas for edge IoT deployment include smart farms with robots and sensors~\cite{farmbeats,ryu2015design,gan2018,chandra21,SMG21}, battlefield deployments~\cite{IoBT,feng2020robustness,shahid2021}, performance arenas~\cite{fox2019live,turchet2018internet}, and even indoor spaces such as warehouses~\cite{fatima2022production,wang2022iot} and large marketplaces~\cite{lampropoulos2019internet,lin2016human}
with growing markets~\cite{iotagrimarket,iotbattlemarket,iotstadiummarket}.

In such physically distributed settings, it is commonplace today to build {\it mesh networks} among the IoT and edge devices. Mesh networks can be robust, resilient, and allow easy ``scale out'' by adding and removing devices, without requiring a central operating unit (or hub) which may be overloaded and prolong latencies. Some devices may be ``smart'' devices capable of compute and storage (both small-scale), while other devices may be ``simple'' devices. Mesh networks
are today being built using Wifi, Zigbee, Bluetooth, LoRa, etc.~\cite{katanbaf21,sayian,curvinglora}. 
Industrial mesh networks using IEEE 802.15.4, WirelessHart~\cite{song2008wirelesshart}, or IEC~\cite{iso_central_secretary_systems_2016} are becoming common. 
There is rich literature at the networking layer, e.g., on how to build, configure, and route inside mesh networks~\cite{akyildiz2005survey,YangWangKravetsWiMesh05}.

Sitting {\it above} the mesh network is the  control plane---this needs to execute three kinds of actions: {\it Sense-Trigger-Actuate} (or STA). The control plane needs to {\it sense}: continuously collect measurements from multiple devices. Applications specify predicates, which are arbitrary Boolean clauses involving multiple device  readings. When a predicate is satisfied, the control plane  needs to {\it trigger} programmed series of actions that touch many devices. Third, because some of these actions may be long-running (e.g., mechanical movements, repetitive actions, or time-based actions such as water sprinklers), the control plane needs to initiate and monitor {\it actuation} of these commands (at multiple devices), while satisfying the challenging goal of maintaining consistency~\cite{Safehome}.

At a smaller scale, this STA abstraction exists in smart homes. 
The most popular programmatic abstraction for STA in smart homes, which is also applicable in edge meshes, is a {\it routine}. A routine is a sequence of {\it commands} that touches multiple devices, wherein each command executes an action on only one device. Multiple routines may be active at any given time, while many more  remain  dormant, waiting to be triggered. A routine may be triggered either based on sensor values, or at particular times, (or a combination thereof), or manually.  Routines are offered today by Amazon Alexa~\cite{Alexa}, Google Home~\cite{GoogleHome}, Samsung SmartThings~\cite{SmartThings}, and others. For instance, one routine at an office building may be triggered every week day at 9 pm, switching on security cameras, switching on specific external lights, and locking the outside doors. Another routine might be triggered by a combination of sensors and time, e.g., if the building temperature drops below 45 F between 11 pm and 6 am, take emergency actions to prevent bursting of pipes, and send  notifications to facilities staff. During the day multiple routines may run continuously, e.g., some raising and lowering shades on the exterior of the building in order to regulate sunlight flowing into the building as a function of temperature, while other routines change interior lighting in different rooms and corridors as a function of occupancy, etc.  Routines may be triggered either by humans (on demand) or automatically.

In a smart home the small number of devices, and the physically proximate nature of the home, means that a single central hub (e.g., an Alexa unit or a Google Home unit) can be used to execute all the three S-T-A categories of actions in the control plane. However, in   commercial IoT deployments, the central hub approach becomes untenable as these deployments 
span large areas, involve many tens or hundreds of devices or more, and routines (active+dormant) numbering in several tens to a few hundred. In fact, a centralized hub may even be infeasible to even set up in really large areas like smart farms and battlefields. Hence mesh networks are growing in popularity for such settings. 

Where it is possible to set up, the central hub may be overwhelmed very quickly by S-T-A traffic. For the setting with routines that we described, with a (say) 1 second fidelity for monitoring devices that trigger routines,  monitoring just 500 devices in a building involves the  home hub exchanging messages every 2 ms; even using AWS IoT's liberal (suggested) 30 s ping frequency \cite{AWSIoT} means the central hub exchanges messages every 40 ms.

Distributed support for STA  needs to be {\it local}, without needing a connection to the cloud. Central hubs in use today are themselves unreliable~\cite{moore2020iot}, and almost always connect to the cloud which itself may suffer outages~\cite{gcloud-down1}.  No mechanisms exist today to coordinate  multiple local hubs for the specific STA workload  of routines running over large edge meshes. 

This paper presents \sysname{} ({\it Control Mesh}),
the first system for  monitoring and management of routines (like those available via Amazon Alexa, Google Home, Samsung SmartThings, etc.) over edge meshes.
\sysname{} is naturally  fully-decentralized  and contains {\it local} distributed protocols running among available smart devices (other simple devices unable to run compute or memory, may also co-exist)---only a small fraction of devices need to be smart for \sysname{} to work. \sysname{} relies on neither a cloud nor on a local central hub.

\sysname{} has to tackle four major challenges: (1) {\it fully-decentralized} monitoring, triggering, and actuation of routines, (2) {\it load-balance}  work across both devices and  time, (3) {\it tolerate simultaneous failures} of multiple devices, and (4) have load that is {\it aware of the physical layout of devices}. We propose a building block called a {\it k-group}, which is a dynamically selected subgroup of smart devices. 
This allows us to break the STA workload into fine granularities, by assigning a separate k-group for sensing each device (or a group thereof), and one k-group for triggering and actuating each routine. \sysname's k-groups use a {\it combination of spatial load balancing and temporal load balancing}: a) each k-group is selected via consistent hashing, and b) migrated periodically. The migration spreads the load of ``heavy'' subgroups (e.g., k-groups assigned to devices whose readings vary significantly, or k-groups assigned to frequently triggered routines) over the system. Within each k-group, we eschew using internet-based coordination systems like Zookeeper~\cite{zookeeper} which may be expensive, and instead we carefully design fast and cheap versions of coordination techniques including agreement via quorums, leader election, and responding to failures of subgroup leader or members. Specifically, to tolerate $f$ failures, it suffices for each subgroup to contain $(2f+1)$ members. 

To make proactive k-group migrations cheap, \sysname{} uses {\it zero-message exchange} protocols---``zero-message'' means that any node can (in the fast path), without exchanging any messages, calculate the current subgroup for any other given device or routine. In large spaces it is important keep communication  local for latency and energy reasons---\sysname{} adapts ideas from the space of {\it locality-sensitive hashing (LSH)}~\cite{datar_locality-sensitive_2004} to IoT and smart space settings, so that sub-group selections are in the locality of the monitored device(s), and yet retain the benefits of zero-message exchange mechanisms.

Underneath \sysname{} we require: i) any mesh routing algorithm, and ii) a weakly consistent membership protocol
for mesh networks (e.g.,  Medley~\cite{medley}).

The contributions of this paper are: 
\begin{itemize}
    \item We present \sysname{}, the first local and fully-decentralized system for routine monitoring and management over edge meshes intended for sense-trigger-actuate scenarios. 
    \item \sysname{} proposes a building block called k-groups, which help load-balance sensing, triggering and actuation work over devices. k-groups are selected using zero-message mechanisms and locality sensitive hashing, and within k-groups are running fast and cheap protocols for coordination and election.  
    \item We analyze and theoretically prove safety, liveness, and progress of \sysname's mechanisms.
    \item We perform large-scale simulations as well as smaller-scale Raspberry Pi-4 deployment experiments. 
\end{itemize}
\section{System Model}
\label{sec:system-model}

We assume a crash-recovery model: IoT devices may fail (crash) at any time, and then recover at a later time. We do not target Byzantine (malicious) and leave them to future work. As is traditional in fault-tolerant literature, we assume that only up to $f$ simultaneous crash failures occur in the system. The value of $f$ can be calculated from the deployment settings. For instance, buildings contain power domains, each connected to a power breaker.
A common failure mode is an entire power domain going out, thus making all the devices inside that domain fail. Thus one rule of thumb is
to set $f$ to the number of smart devices in the {\it largest} power domain  inside the building. Because a building contains many power domains, $f$ remains much smaller than $N$, the total number of devices in the building.

We assume a subset of devices ($> 3$) are smart devices, equipped with (small) memory and computation capability---other simple (non-smart) IoT devices may also co-exist in the system. Smart devices may include any combination of IoT devices capable of arbitrary compute, Raspberry Pis, home hubs, handheld devices, etc. We assume non-smart devices can only receive and process commands, and respond to queries for status, but they cannot run any computation.

Since IoT devices are typically fixed early in the operation of a building, we assume that the locations of all devices are known by the \smart{}s~\footnote{Device motion can be treated as a failure event, followed by recovery event at its new location.}.  We  assume \smart{}s' clocks are synchronized (although \sysname{} works with loose synchronization).

\begin{figure}[t!]
    \centering
    \includegraphics[width=0.45 \textwidth]{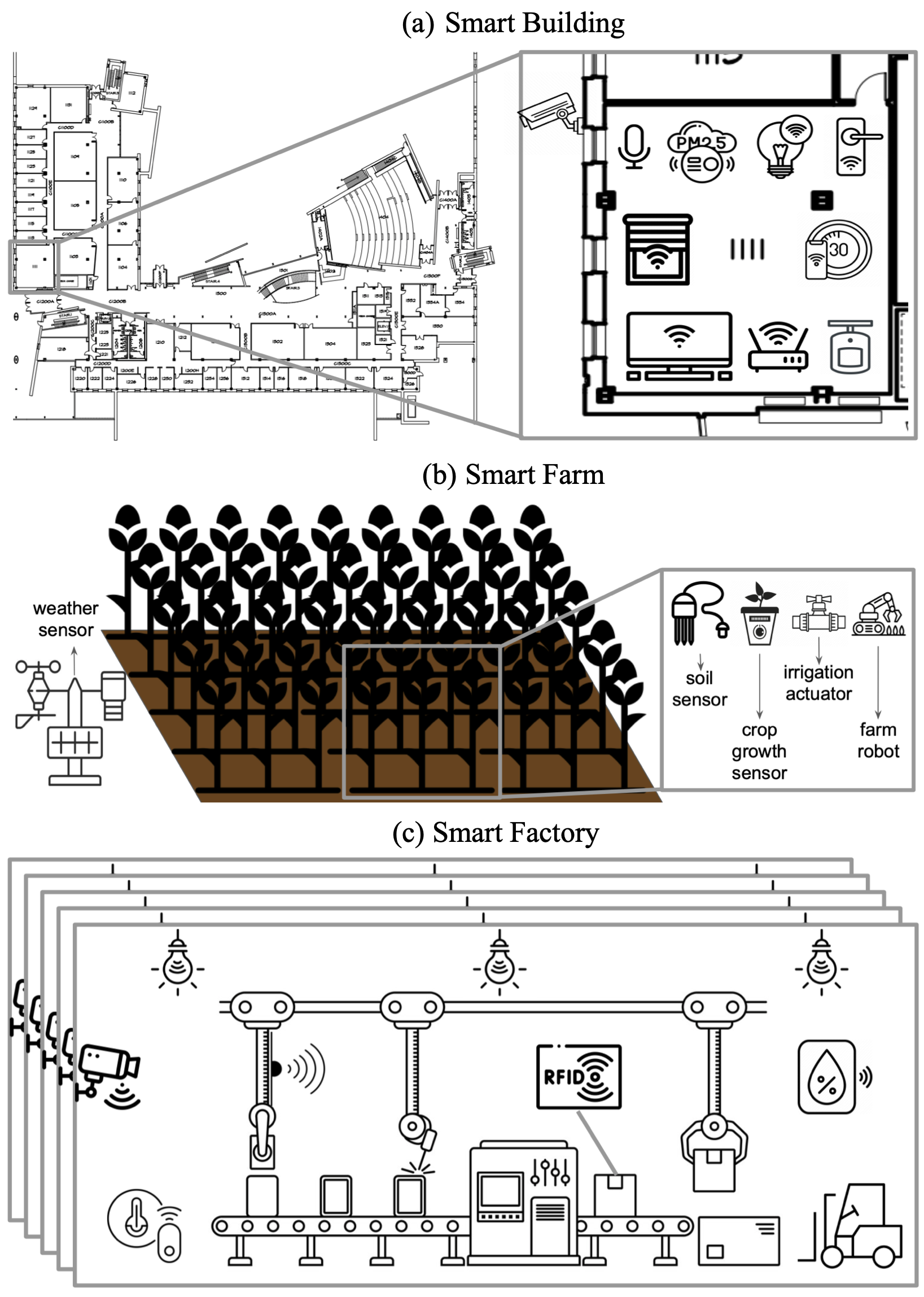}
    \caption{\it \small Three Sense-Trigger-Actuate Deployment Settings: (a) Smart Building, (b) Smart Farm, (c) Smart Factory.}
    \label{fig:IoTsetting}
\end{figure}

Smart devices communicate via a wireless  network that is lossy. For generality, we assume the existence of a wireless ad-hoc routing protocol among smart devices \cite{dijkstra, aodv, dsr}. 
We assume any smart device can send a command to any IoT device.  Finally, \sysname{} is built atop a weakly-consistent membership service like Medley \cite{medley}, which helps maintain a full membership list (of all IoT device IDs that are alive) at each \smart{}. Medley provides $O(1)$ detection time of failure and recovery events. 
Table~\ref{tab:terms} summarizes key terms, and \Figure~\ref{fig:IoTsetting} shows an example setting.

\begin{table}[t!]
    \centering
    \renewcommand\arraystretch{1.3}
    \begin{tabular}{  m{1.8cm} | m{6cm}  }
        \hline
        \strut \small \textbf{Term} & \small \textbf{Description} \\
        \hline\hline
        Command & A message sent to, and associated action executed by, one IoT device.  \\
        \hline
        Routine & A sequence of commands. Triggered either by time, or a set of triggering clauses or manually. \\
        \hline
        IoT Device & All IoT devices in system. \\
        \hline
        Smart Device & An IoT device that has (small) memory and computation capability.  \\
        \hline
        Simple Device & An IoT Device that is not a Smart Device. Typically contain sensor(s) or actuator(s).
 \\
 \hline\hline
    \end{tabular}
    \vspace*{0.2cm}
    \caption{\it \small Key Terms used in this  paper.}
    \label{tab:terms}
\end{table}

\section{\sysname's subgroups: The \kgroups{} Mechanism}
\label{section:design}

\sysname{}'s protocols rely on its  subgroup formation and maintenance. \sysname{} uses these subgroups to sense-trigger-actuate routines and devices (described in  following sections). We describe subgroup formation in this section. 

\sysname{}'s subgroups are called {\it k-groups}, signifying our goal of keeping a \kgroup{}'s size to be around (a small value) $k$. $k$ is a globally configured variable. To tolerate $f$ system-wide failures, we set $k = (2f+1)$.
We present  \kgroup{} selection (\Section~\ref{subsection:member_selection}), election (\Section~\ref{subsection:leadr_elctn}),  quorums (\Section~\ref{subsection:quorum}), temporal migration (\Section~\ref{subsection:periodicity}), failure response  (\Section~\ref{subsection:fault-tolerance}), and locality  (\Section~\ref{subsection:lsh}).

\subsection{\kgroup{} Member Selection}
\label{subsection:member_selection}

 \sysname's \kgroups{} need to be selected in way that is: (i) load-balanced, 
 and (ii) uses a minimum number of messages. 

\noindent \textbf{Load-balanced Selection via Zero-message Exchange}: A smart device with ID $S$ discovers if it is present in a \kgroup{} for a target device with ID $D$ by calculating a consistent hash: $Hash (S, D)$. The {\it lowest} $k$ hashed IDs across the entire system are then the approved \kgroup{} members. This design is inspired by the idea of cryptographic sortition~\cite{micali1999verifiable, cryptosort}. 
Notice that, given a fixed hash  function (e.g., a cryptographic hash like SHA-2 or MD-5), {\it any} arbitrary smart device can calculate the {\it entire} membership of any \kgroup{} by calculating this hash function on each of the IDs from its (full) membership list. With correct membership lists, this takes zero messages. 

\subsection{Leader Election}
\label{subsection:leadr_elctn}
The \kgroup{}'s goal is to fault-tolerantly replicate the state of its monitored entity. To make its operations efficient, each \kgroup{} elects a leader. The alive smart device with the lowest hashed value ($Hash (S, D)$) is considered to be the leader. While election could be done via a zero-exchange mechanism, in order to ensure that the leader is actually alive, \sysname{} runs the well-known Bully election algorithm to quickly elect the leader~\cite{Bully}. The Bully algorithm is attractive as it is fast in the common case: 1 RTT for the lowest-hashed node (leader) to inform others in the \kgroup{}. % that it is the leader. 
If failures occur, the Bully algorithm takes a worst case of 5 RTTs.

\subsection{Quorum Agreement}
\label{subsection:quorum}

Fast agreement in a \kgroup{} is done via quorums. In a \kgroup{} of size $k=2\f+1$, anyone in a group (typically the leader) sends a message to all $M$ members, and waits for acknowledgments from at least $(\f+1)$ members. Thus with $\f$ failures, at least one surviving \kgroup{} member knows about the latest decisions in the \kgroup{}, and can communicate it to future leaders.

\subsection{Epoch-based \kgroup{} Migration}
\label{subsection:periodicity}

While Section~\ref{subsection:member_selection} is {\it spatially} load-balanced, for temporal load balancing, \sysname{}  migrates each \kgroup{} membership periodically. This is done once every {\it epoch}, and epoch lengths are fixed system-wide. (Different \kgroups can migrate at different times.)

First we extend the hash function for the selection to include the epoch number, making it $Hash (epoch, S, D)$. The $epoch$ is  incremented by 1 on each epoch change. Next, when an epoch expires, the following sequence executes: 1) the new \kgroup{} is formed by using {\it Hash (++epoch, S, D)}; 2) an election is run in the new \kgroup{} (using \Section{}~\ref{subsection:leadr_elctn});  3) any state maintained by the leader in the old \kgroup{} is migrated to the new \kgroup{}'s leader;  4) the new \kgroup{}'s leader replicates the state information at a quorum of its new \kgroup{} members (\Section{}~\ref{subsection:quorum});  5) {\it atomic changeover to new \kgroup{}:} new \kgroup{} leader tells old \kgroup{} leader that the old \kgroup{} is decommissioned, old leader acknowledges; and 6) old leader tells old \kgroup{} members that they are decommissioned. 

When the new \kgroup{} leader acknowledges to the old \kgroup{} leader in step 5 that it has received all the state, the old \kgroup{} leader can delete its old \kgroup{} state. After step 6, an old \kgroup{} node can delete its old state for that \kgroup{}. During steps 1 though 7, no monitoring operations occur in the \kgroup{}. This means that old \kgroup{} members and old leader cease all \kgroup{} operations at exactly the epoch changeover boundary.  In practice the epoch change is fast enough that this gap in monitoring (between steps 1 to 6) is small, as shown by our experimental results. After step 6, all normal \kgroup{} operations resume.

\subsection{Failure Response: \kgroup{} Updates}
\label{subsection:fault-tolerance}

If a smart node $N_i$ fails, the failure detector layer informs all other nodes of its failure. Because \sysname{}'s operations remain correct for group size $ \in [f+1, 2f+1]$, we trigger a \kgroup{} selection only when $f$ nodes have failed (rather than on each failures)---this keeps the overhead of failure response low.

The only exception is if the failed node $N_i$ was a leader of a \kgroup{}. This causes Section~\ref{subsection:leadr_elctn}'s election algorithm to be re-run. 
This new leader needs to reconstruct the state from all the nodes in the \kgroup{}. Each piece of state needs to be confirmed by a quorum ($f+1$ nodes). This state is correct because all past updates were present at at least $f+1$ nodes, so the new leader receives each past update from at least one non-faulty node.
If the new leader fails during reconstruction, election is re-run and the process repeats.

\subsection{Grouping via Locality-Sensitive Hashing}
\label{subsection:lsh}

Because the underlying routing infrastructure is an ad-hoc network, this raises two issues. 
First, the sheer number of devices and routines may lead to a large number of \kgroup{}s. 
Second, a randomly selected \kgroup{} will consist of nodes that are spread across the ad-hoc network, thus incurring high message overhead on links.

\noindent {\bf Limiting the Number of \kgroup{}s:} We use a %$k$-means 
clustering algorithm \cite{KMeans}  to partition the devices into clusters. Each cluster is assigned to one \kgroup, with a random device from the cluster chosen as the ``center''  (i.e., representative) in order to calculate the locality sensitive hashing function described above. For routines, we pick a random device it touches as its representative, and that device's cluster is now the routine's cluster (and thus its \kgroup). 

\noindent {\bf Locality-Sensitive Hashing: } Selecting \kgroup{}s randomly (Section~\ref{subsection:member_selection}) spreads the \kgroup{} members all over the network, slowing down intra-group operations (election, quorum, inter-leader).
The challenge is to make \kgroup{} selection {\it local} while still retaining many of the spatial load-balancing benefits of hashing. To do so, we need to adapt ideas from locality-sensitive hashing (LSH)  \cite{datar_locality-sensitive_2004}, an idea from high-dimensional data science. To the best of our knowledge, our paper is the first to apply LSH to edge/IoT networks. 

Concretely, we modify the \kgroup{} member selection to be a two stage process: select $f+1$ nodes in the vicinity of the monitored device/cluster, and an additional $f$ nodes randomly from across the group. This provides for speed in intra-group operations (via nearby $f+1$ members), while still ensuring that a power domain failure (see Section~\ref{sec:system-model}) does not compromise the operation of a group (since there will be fewer than $f+1$ devices in a power domain). 

The main idea of LSH~\cite{datar_locality-sensitive_2004} is to hash similar vectors (in space) into the same ``buckets" with high probability. A randomly selected ``center" (from the monitored device group) and the smart devices are hashed into buckets and the majority is selected from the smart devices that are in the same buckets with the center ``candidates"). If there are more than $f+1$ candidates, we choose the nearest $f+1$ ones to the center. Notice that selecting the majority randomly from the candidates usually does not work well since buckets may also have nodes that are not close to the center. If there are fewer than $f+1$ candidates, \sysname{} first checks if the center's neighbors (devices that are close to the center) have any candidates that we can ``borrow", before moving to random selection. 

Overlapping of \kgroup's members in consecutive epochs helps with data transfer operations during epoch change.  
To achieve this, we append $(pre_i, \dots, pre_d')$ to its own location vector $(x_1, \dots, x_d')$, where $pre_i$ is the average  location of all devices this smart device monitored during the previous epoch on $i$th dimension. Then we use this vector as an input for LSH. Adding some randomness to the vector entries for load balance purposes is also an option (our default setting). We append $(x_1, \dots, x_d')$ to a simple device's location vector to make all devices have the same dimensions.

The resultant locality-sensitive hash function $h_{\bm{a},b} = \lfloor{ \frac{\bm{a\cdot v}+b}{r}} \rfloor$ adapted from  \cite{datar_locality-sensitive_2004}, maps a $d$ dimensional vector $\bm{v}$ onto a set of integers, where $a$ is a random $d$ dimensional vector whose entries are independently chosen from a $p$-stable distribution (Gaussian distribution in our case) \cite{zolotarev1986onedimensional,datar_locality-sensitive_2004}; $b$ is a real number uniformly chosen from $[0,r]$ and $r$ is a parameter. To further reduce the probability that vectors with very different entries fall into the same buckets, several individual hash functions can be concatenated, such that $H = (h_1, ..., h_m)$). Additionally, we can use $l$ hashes ($H$s) to increase the number of similar vectors in the same buckets. That is, we have $\mathcal{H} = \{H_1,..., H_l\} =\{(h_{11}, ..., h_{1m}),..., (h_{l1}, ..., h_{lm})\}$.

\begin{figure}[tb!]
    \centering
    \begin{subfigure}[b]{0.48\linewidth}
        \centering
        \includegraphics[width=1.1\textwidth]{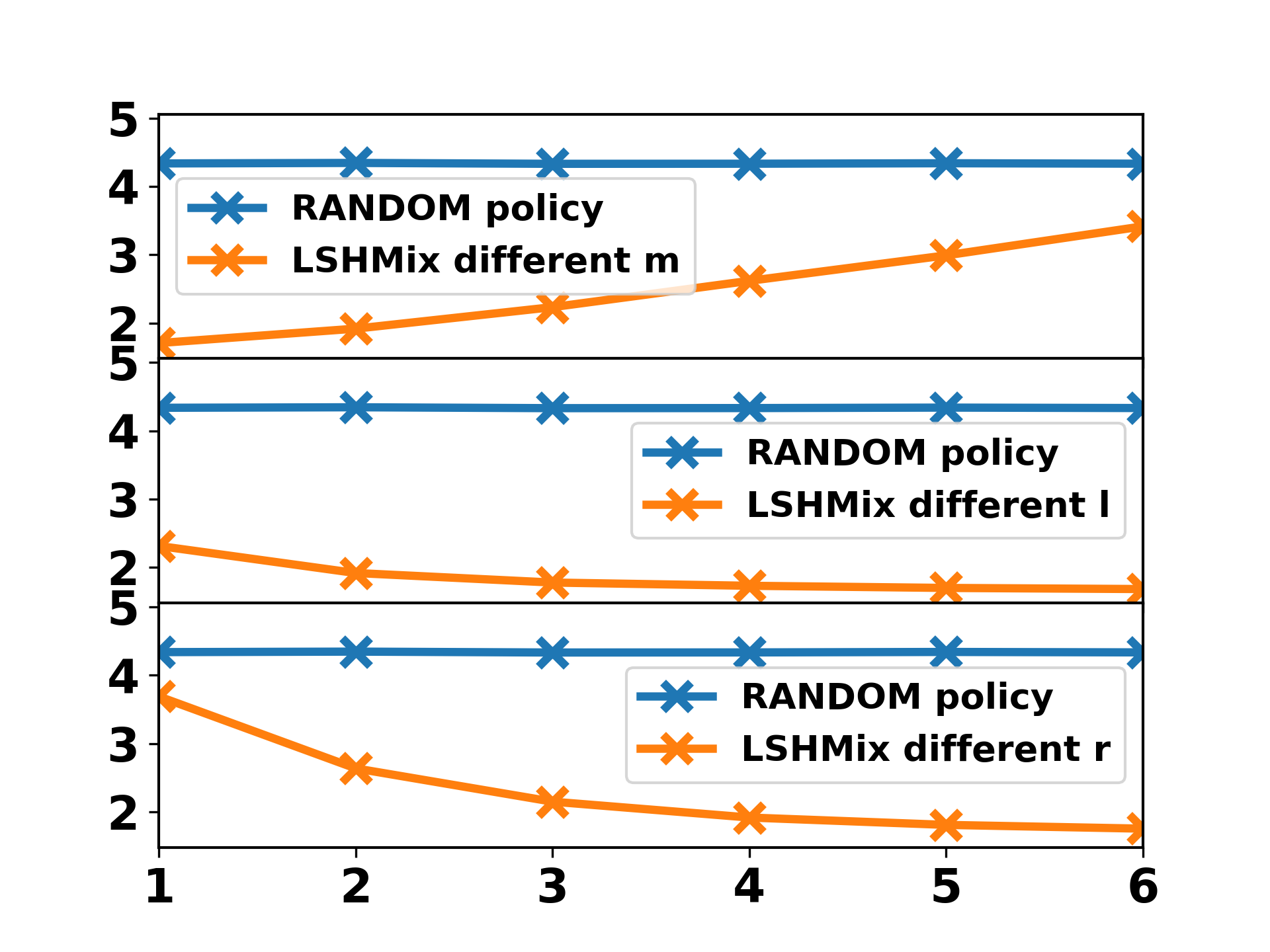}
        \vspace{-12pt}
        \caption{\small \it Average quorum distance. $N$ = 500, \#smart devices = 200.}
        \label{subfig:quorum_dis_500klr}
    \end{subfigure}
    \hfill
    \begin{subfigure}[b]{0.48\linewidth}
        \centering
        \includegraphics[width=1.1\textwidth]{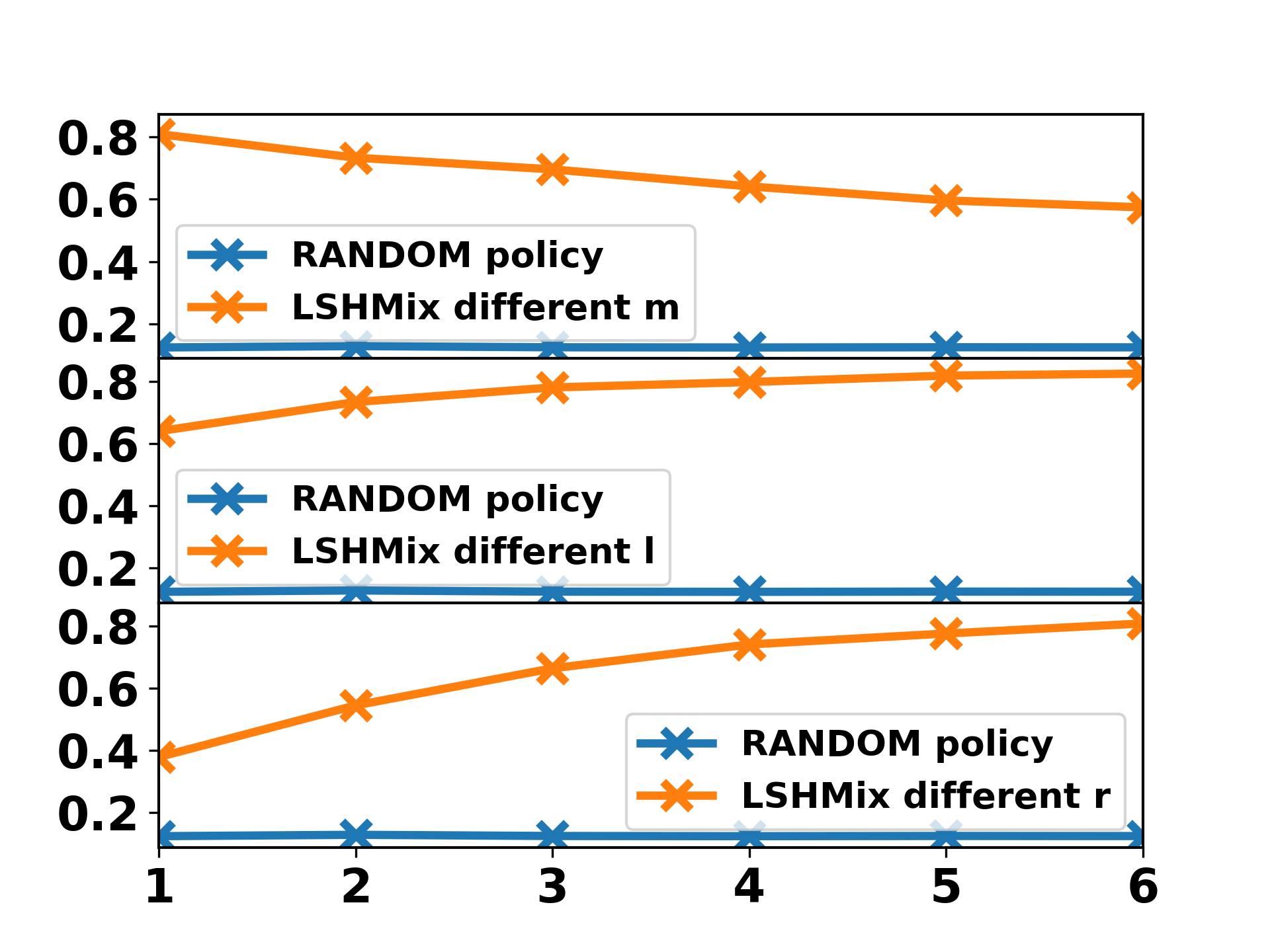}
        \vspace{-12pt}
        \caption{\small \it Average \kgroup{} overlap. $N$ = 500, \#smart devices = 200.}
        \label{subfig:overlap_500klr}
    \end{subfigure}
    \par\medskip

    \begin{subfigure}[b]{0.48\linewidth}
        \centering
        \includegraphics[width=\textwidth]{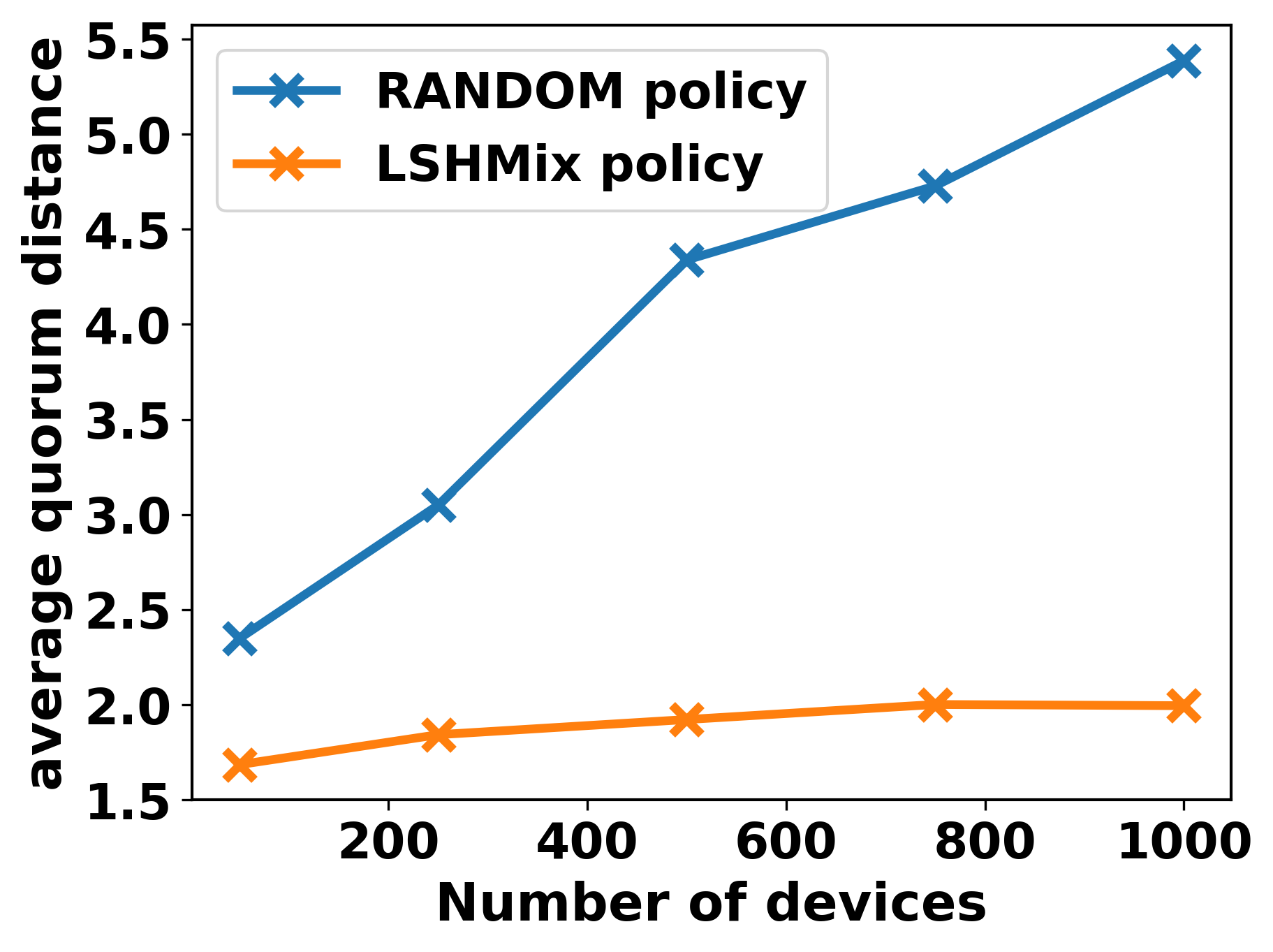}
        \caption{\small \it Average quorum distance for different system sizes (N).}
        \label{subfig:ave_quorum_dis_topology}
    \end{subfigure}
    \hfill
    \begin{subfigure}[b]{0.48\linewidth}
            \centering
            \includegraphics[width=\textwidth]{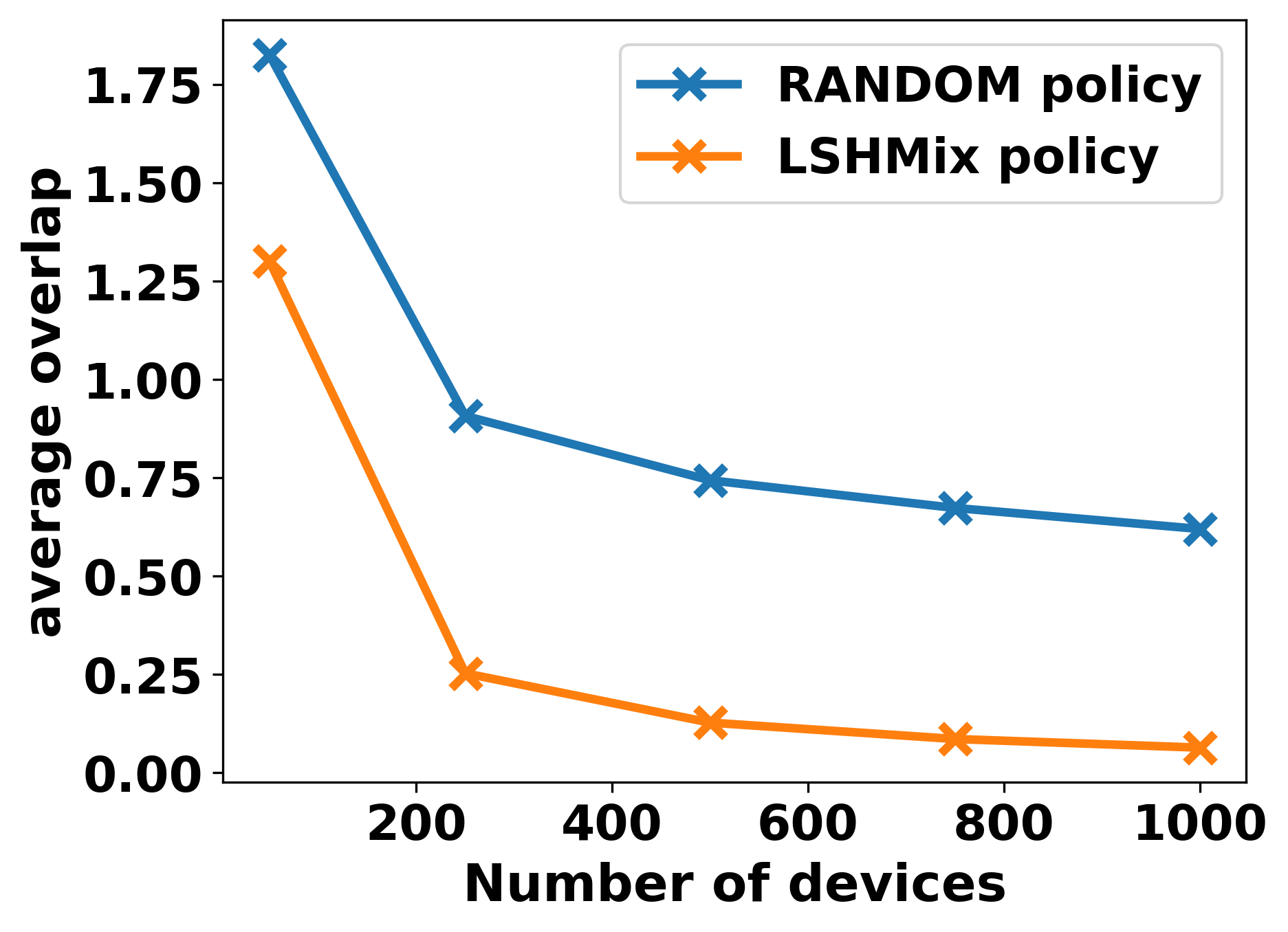}
            \caption{\small \it Average \kgroup{} overlap for different system sizes (N).}
            \label{subfig:ave_overlap_topology}
    \end{subfigure}
    \caption{\small \it Average quorum distance and \kgroup{} overlap for different selection policies, parameters and system sizes.
    } 
    \label{fig: ave_quorum_dis_and_overlap_klr}
\end{figure}

We give some intuition for choosing the values of these LSH parameters. 
Large $m$ values  exclude nodes that are not similar and reduce the number of candidates, while large $l$ and $r$ may increase this number (\Figure~\ref{subfig:quorum_dis_500klr} and \Figure~\ref{subfig:overlap_500klr}). The default setting in our paper is $m=2, l=2, r=4$.
In \Figure~\ref{subfig:ave_quorum_dis_topology} and \Figure~\ref{subfig:ave_overlap_topology} we tested our default setting for grid topologies with different sizes.
We can see that while the average hop count to achieve quorum (quorum distance) increases linearly with the system size for the random selection policy, it stays small for the LSH-based selection policy. More experiments on random and clustered networks can be found in Sec.~\ref{section:exp_eval}.
\begin{table}[tb!]
    \centering
    \renewcommand\arraystretch{1.3}
    \begin{tabular}{m{1.5cm} | m{0.95cm}| m{0.95cm}| m{0.95cm}|m{0.95cm} | m{0.95cm}}
        \hline
        $N$ & 50 & 250 & 500 & 750 & 1000\\
        \hline
        \#candidates & 13.0223 & 34.1258 & 50.7501 & 52.8035 & 66.9135\\
        \hline
    \end{tabular}
    \vspace*{0.15cm}
    \caption{\small \it Average number of \kgroup{} member candidates for different $N$s. 40\% of the devices are smart ones.}
    \label{tab:num of candidates}
\end{table}

Although the \kgroup{} overlap based on LSH drops for large $N$ (\Figure~\ref{subfig:ave_overlap_topology}), it is much larger than the random policy. We hypothesize that with more nodes, and thus more candidates, the probability of the same device being selected in consecutive epochs grows smaller.
This hypothesis is validated by Table~\ref{tab:num of candidates}. 
One may also adjust LSH parameters to reduce the candidate count without affecting quorum distance. E.g., at $N = 1000$, $m = 3, l = 2, r = 4$, the average number of candidates is 21 and the average quorum distance is 2.4 (vs. 5.3 for the random policy). 

\section{Routine \& Device Management}
\label{section:mngmt}

We now describe how sense-trigger-actuation management of routines and devices builds atop the \kgroups of Section~\ref{section:design}.

\subsection{Device Management}
\label{sec:devmgmt}

In \sysname{} each device is assigned to a unique \kgroup{}. This \kgroup{} is responsible for sensing actions on the device, i.e., tracking the up/down status of the device, and the latest state of the device (e.g., temperature readings for a thermostat). The  leader of the device's \kgroup{} (henceforth called the {\it device leader}) periodically pings the device. If the device does not respond within 2 RTTs, the device is presumed failed, and the device leader notifies all \smart{}s in the topology. If the detected state of the device $D$ (received in the ack) differs from its latest recorded state at the device leader, the device leader sends this new state to all the routines which contain $D$  in their triggering clauses (specifically, sending to the leaders of those routine's respective \kgroups).

\subsection{Routine Management}
\label{sec:routinemgt}
\label{subsection:rtn_mngmt}
Recall that a routine consists of a trigger clause (arbitrary boolean clause involving multiple devices and their states), and a set of commands to execute when the routine is triggered. The set of devices in the former {\it trigger set} (read from) and the latter {\it command set} (actuated on) may be different or may overlap. 

In \sysname{} each routine is assigned a unique \kgroup{}, which is responsible for sensing (monitoring), triggering, and actuation (execution) of the routine. The sensing is done in collaboration with device \kgroups, as described earlier.   
A routine may be in one of the following states (maintained at its \kgroup): 
(i)~\texttt{NOT\_TRIGGERED}, (ii)~\texttt{ACQUIRING\_LOCKS},
(iii)~\texttt{EXECUTING}, and  (iv)~\texttt{RELEASING\_LOCKS}. 
We call the leader node of the routine's \kgroup{} as the {\it routine leader}. The routine leader connects with the device leaders of each device in the routine's trigger set.  Whenever these device leaders  send updated states to the routine leader (Section~\ref{sec:devmgmt}), the latter checks if this state change satisfies the trigger clause for the routine.

When a routine is triggered, the routine leader changes its state  from  \texttt{NOT\_TRIGGERED}, to  \texttt{ACQUIRING\_LOCKS}, and  replicates this state to
all its \kgroup{} members. It waits for  ($\f + 1$) responses, counting  itself. 
When a routine starts it first acquires (virtual) locks to all devices in the routine's command set. Then the routine executes its commands. Finally when the routine is finished it releases all its locks. This pessimistic concurrency control approach has been shown to be fast, avoid deadlocks and livelocks, and scales well~\cite{Safehome}.

The lock for a given device is only maintained in that device \kgroup{} (rather than the routine \kgroup)---this allows multiple routines to compete for locks. For a routine $R$ to acquire a device $D$'s lock, $R$'s routine leader communicates with $D$'s device leader. (If the routine leader is unaware of a current device leader, it contacts all \kgroup{} members of that device, to know the leader via  Section~\ref{subsection:member_selection}'s  zero-message exchange protocol, and then contacts the device leader.) 

The device leader maintains a {\it wait list} of routines requesting a lock, and grants only one lock at a time (in FIFO order). When a lock is released by a routine leader (at \texttt{RELEASING\_LOCKS} stage), the device leader dequeues the next entry from the wait list, sends a request to all its device \kgroup{} members, and waits for $(f+1)$ acks, before it sends back the lock to this requesting routine leader.
When all locks have been acquired, the routine leader changes the routine's state to \texttt{EXECUTING}, communicates this to all routine \kgroup{} members, waits for $(f+1)$ acks,
and then starts issuing the commands for the routine. 
{A routine's command is always routed from the routine leader to the appropriate device leader which then forwards it to the device. This way the device leader (\kgroup) always knows the device's latest status. } 

The order in which locks are acquired can affect the speed of triggering. There are two design options---sequential and optimistic. 

\subsubsection{Sequential Lock-Acquiring (SLA) Strategy}
\label{subsubsec:sla}
The routine leader acquires locks sequentially. If the locks are acquired in increasing order of device ID, this prevents deadlocks (cycles cannot occur) among multiple routines with conflicting device sets. The SLA strategy can be slow when there are no or few conflicts.

\subsubsection{Optimistic Lock-Acquiring (OLA) Strategy}
\label{subsubsec:ola}
 \sysname{}'s Optimistic Lock-Acquiring (OLA) attempts to acquire all locks for its (desired) devices simultaneously. If any lock request fails, the routine leader releases all locks it acquired, and retries all again. To prevent message implosion, retries occur after a backoff timeout. OLA is costly when routines' command sets overlap. 
 
 We experimentally compare these two strategies later.

\section{Formal Analysis}
\label{section:analysis}

We formally analyze \sysname{}'s properties. 
For brevity, we intuitively summarize our findings first. Then we state the formal theorems. The proofs can be found in Appendix A.

\squishlist
\item {\it Inheritance}: When a k-group changes, the state held by the old leader and new leader are identical.
\item {\it Safety}: No two routines that touch an overlapping set of devices, are allowed to execute simultaneously. 
\item {\it Liveness}: No routines deadlock.
\item {\it Progress: Every routine makes progress.}
\item We also calculate the (probabilistic) availability of the system when more than $f$ devices fail. 
\squishend

\begin{definition}
Inheritance means that given a \kgroup{}, the old and new versions of the \kgroup{}---before and after epoch change, or after leader failure---maintain identical state. 
\end{definition}

We now state the key results.

\begin{lemma}
\label{lemma:inheritance}
{\bf [Inheritance] }Inheritance is guaranteed after the state transfer stage when at most one node fails.
That is, the state held by an old \kgroup{} leader is the same as the state held by
the new \kgroup{} leader after the state transfer stage, under failure of at most one node.
\end{lemma}

\label{subsec:safety}

\begin{theorem}
\label{theorem:safety} {\bf [Safety]} 
No two routines that conflict in devices (i.e., their touched device sets are not disjoint) can execute simultaneously, when there is at most one node failure. This is true for both SLA and OLA locking strategies  (Section~\ref{subsubsec:sla}).
\end{theorem}

\label{subsec:liveness}

\begin{theorem}
\label{theorem:deadlock} {\bf [No Deadlocks]} 
No two routines are stuck in a deadlock.
\end{theorem}

\begin{theorem}
\label{theorem:progress} {\bf [Progress/Liveness]} 
Consider a set of executing routines $\mathcal{R}$ each %currently
 still waiting to acquire 1 or more locks, and the (union) set of devices $D$ they are waiting for. If no further routines arrive into  $\mathcal{R}$, and SLA locking is used,  then: at least one routine from the set $\mathcal{R}$ will make progress (i.e., get access to one more device that it desires).
\end{theorem}

\begin{figure}[t]
    \centering
    \begin{subfigure}[b]{0.48\linewidth}
        \centering
        \includegraphics[width=\textwidth]{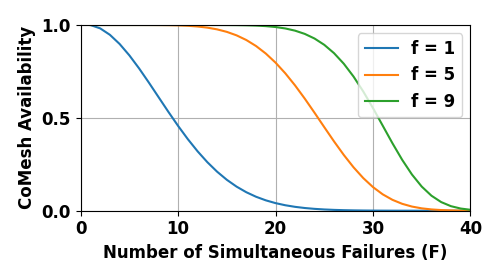}
        \vspace{-6pt}
        \caption{\small \it Availability under different number of simultaneous failures.}
        \label{subfig:prob-on-f}
    \end{subfigure}
    \hfill
    \begin{subfigure}[b]{0.48\linewidth}
        \centering
        \includegraphics[width=\textwidth]{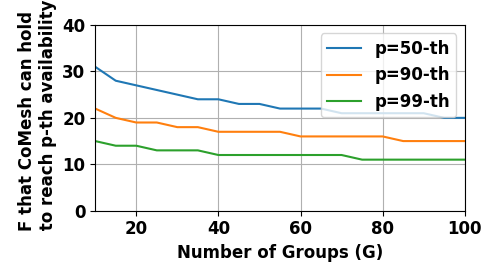}
        \vspace{-6pt}
        \caption{\small \it Impact of different number of groups in the system}
        \label{subfig:prob-varyg}
    \end{subfigure}
    \vspace{6pt}
    \caption{\small \it Probability of \sysname{} running correctly under different parameters. (Default S=100, G=30, f=5)
    } 
    \label{fig:prob-analysis}
\end{figure}

\begin{theorem}
\label{theorem:availability} {\bf [Availability]} 
With $F (>f)$ simultaneous failures, the system remains available, i.e., all \kgroups{} contain at least $(f+1)$ non-faulty nodes, with probability $P(\F{})$ that varies as:  

\begin{enumerate}
    \item When $\F{} \le \f{}$, $P(F) = 100\%$
    \item When $\F{} \ge \f{}$, the {\it availability} is: 
        \begin{align*}
        \label{equ:pf}
          P(F) &= P(\text{at most } f \text{ failures across } k \text{-groups}) \\
               &= P(\text{at most } f \text{ failures in one } k \text{-group})^G  \\
               &= \Bigg(\sum_{i=\max(0, k+F-S)}^{f}\frac{\binom{S-F}{k-i}\times \binom{F}{i}}{\binom{S}{k}}\Bigg)^G
        \end{align*} 
\end{enumerate}

\end{theorem}

{Practically, this scales.  \Figure~\ref{subfig:prob-on-f} shows that as $F$ rises, \sysname{} availability drops as expected but still stays above 50\%  when $\F{}$ grows up to 9$\times f$.  \Figure~\ref{subfig:prob-varyg} shows that in order to reach high availability ($p=50\%,90\%,99\%$), the required $F$ value drops very slow---this indicates scalability with the number of groups.
}

\section{Simulation Evaluation}
\label{section:exp_eval}

To perform large-scale experiments with 100s of nodes on \sysname, as well as to measure  its internal behavior, we wrote a custom simulator. Higher-fidelity simulators like NS3  have known scalability issues beyond a couple of hundred nodes. We measure both: i) user-facing metrics, e.g., latency to start routines, load balancing, etc., and ii) microbenchmarks of \kgroup{} operations. 

Our simulator models network contention by scaling down  network bandwidth based on density: specifically  sending bandwidth from node $A$ to node $B$ is set to the default node bandwidth divided by  number of 1-hop neighbors of $A$. We set  default node bandwidth (Table~\ref{tab:def_param_vals}) conservatively based on bandwidth numbers reported via user complaints  on online forums about IoT devices~\cite{google-nest-speed, rpi-slow-speed1}.
Latency on each hop is fixed to 1 time unit, as we assume that transmission radii are small. Ad-hoc routing is based on Dijkstra's shortest paths.

\begin{table}[tb!]
    \centering
    {\small
    \renewcommand{\arraystretch}{1.2}
    \begin{tabular}{r|c|l}
        \strut \small \textbf{Parameter} & \small \textbf{Notation} & \small \textbf{Default value} \\
        \hline \hline
        \# all devices & $N$ & 250 \\
        \# failures for completeness & $\f$ & 2 \\
        k-group size & $K$ & 5 \\
        \# smart devices & $S$ & 100 (40\% of all devices) \\
        \#seeds & -- & 10 \\
        epoch length & -- & 200 \\
        max routine length & -- & 5 \\
        avg devices per routine & -- & 5 \\
        \# device managed by \kgroup{} & -- & 25 \\
        bandwidth cap & -- & 625 KBps/5Mbps \\
        \kgroup{} selection policy & -- & LSH-random mix \\
        device cluster policy & -- & locality \\
        leader election policy & -- & LSH smallest hash\\
        default LSH parameters & -- & m = 2, l = 2, r = 4\\
        \hline
    \end{tabular}
    }
    \vspace*{0.15cm}
    \caption{\small \it Default parameter values}
    \label{tab:def_param_vals}
\end{table}

We use three different network topologies: 3D grid, random, and clustered. %For  clustered,   
We ensure the graph is connected. 
\sysname{} was configured with topology-aware LSHMix (\Section~\ref{subsection:lsh}). \Table~\ref{tab:def_param_vals} shows default parameters.

\subsection{Client Delay}

\begin{figure}[tb!]
    \centering
    \begin{subfigure}[b]{\linewidth}
        \centering
        \includegraphics[width=0.75\linewidth]{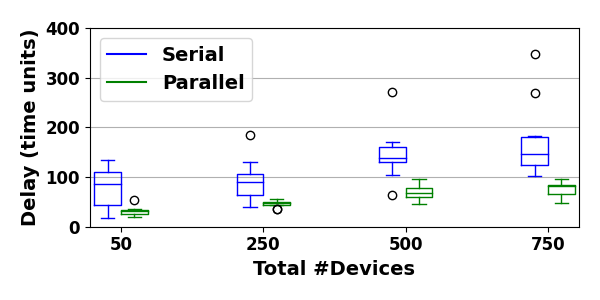}
        \vspace{-6pt}
        \caption{\small \textit{Grid} topology}
        \label{subfig:client_delay_grid}
    \end{subfigure}
    \par\medskip
    \begin{subfigure}[b]{\linewidth}
        \centering
        \includegraphics[width=0.75\linewidth]{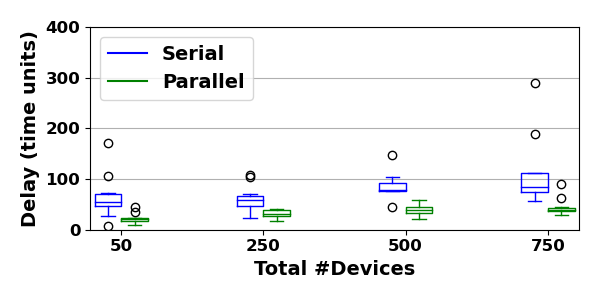}
        \vspace{-6pt}
        \caption{\small \textit{Random} topology}
        \label{subfig:client_delay_random}
    \end{subfigure}
    \par\medskip
    \begin{subfigure}[b]{\linewidth}
        \centering
        \includegraphics[width=0.75\linewidth]{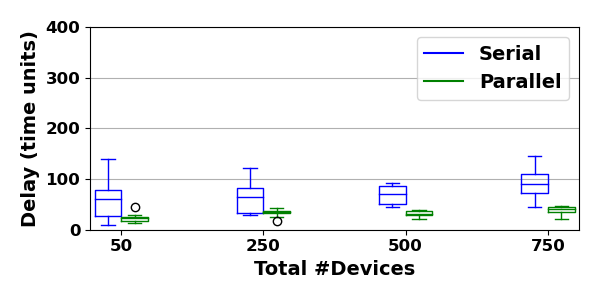}
        \vspace{-6pt}
        \caption{\small \textit{Clustered} topology}
        \label{subfig:client_delay_cluster}
    \end{subfigure}
    
    \caption{\small \it Client Delay (sys) Under Serial vs. Parallel Locking Strategy in three different topology setups 
    }
    \label{fig:pl-sr-comp}
\end{figure}

When a routine's trigger conditions become true, how long does it take for the routine to execute its first command, given there are no other routines active? This is the {\it client delay} for a routine: the time it takes for its \kgroup{} to
detect (and replicate) the trigger, lock all of its touched devices, and start execution. 
In \Figure~\ref{fig:pl-sr-comp}, we observe that \sysname{}'s client delay scales with   device count, regardless of topology or locking strategy (SLA = serial or OLA = parallel). The trend is gradually increasing in all plots. 

As expected the parallel locking (OLA) strategy is faster than serial locking (SLA) because client delay excludes  conflicting routines. Among the 3 topologies, grid  has higher client delay than random and clustered (which are similar) because grid topology's devices are more spread out. {That is, while all three topologies have the same overall density, the variance of density in different parts of the network is higher for random and clustered topologies, causing ``clusters'' of nodes to form and speeding up communication. } 

\subsection{Churn Effect on Client/Sync Delays}

\begin{figure}[tb!]
    \centering
    \includegraphics[width=0.45 \textwidth]{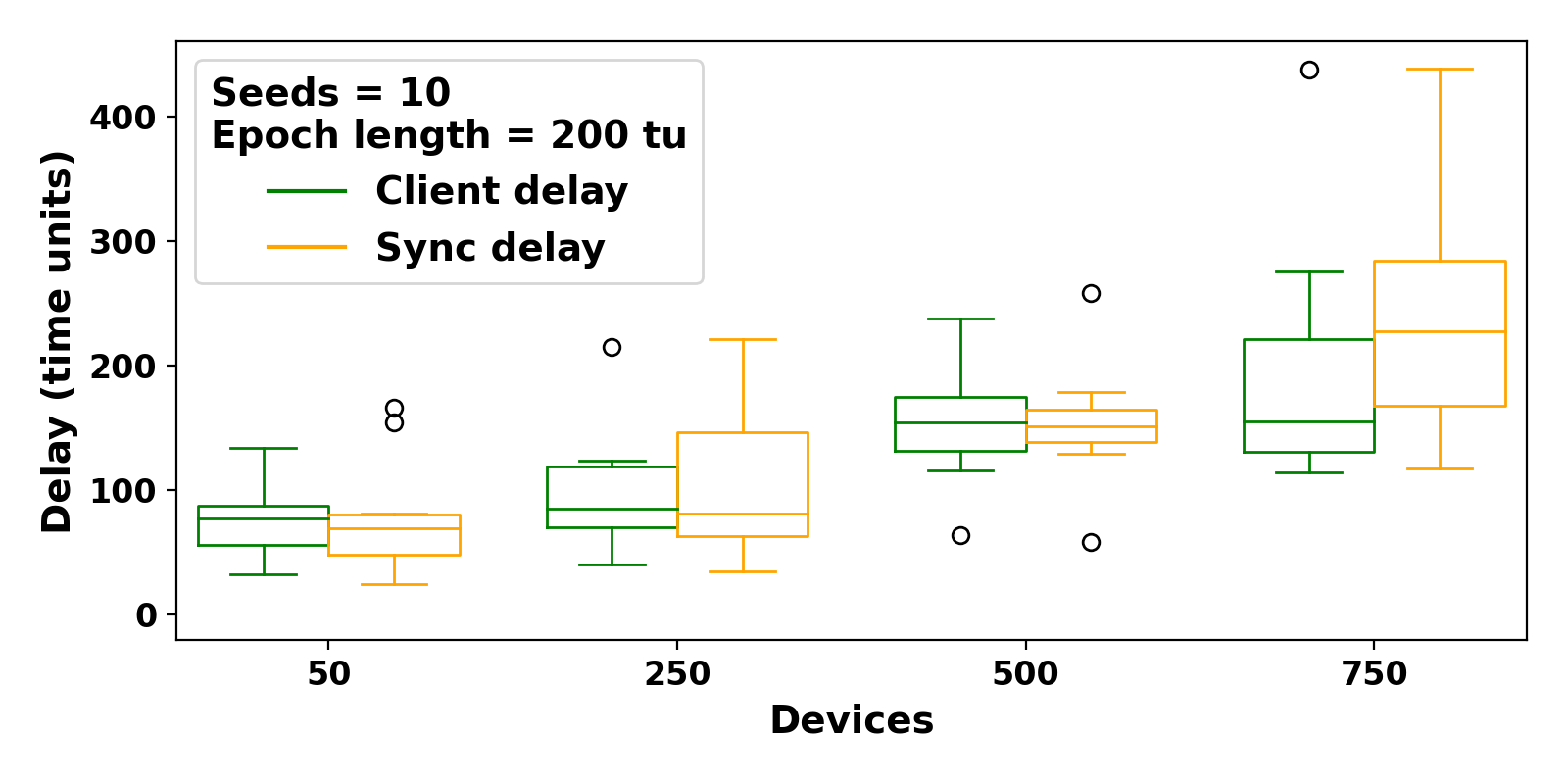}
    \vspace{-6pt}
    \caption{\small \it Client Delay %(sys) 
    and Synchronization Delay when 40\% of the \smart{}s in a grid topology experience churn (either a join or a failure)}
    \label{fig:client_sync_delay_failures}
\end{figure}

\Figure~\ref{fig:client_sync_delay_failures} shows that when 40\% of the \smart{}s' nodes are churned 
in a grid topology (SLA locking), the client delay  increases only 25\%.
This extra delay accounts for the leader election and state transfer to the new leader after
a \kgroup{}'s old leader has failed. 
We define {\it synchronization delay} between a routine releasing its last  conflicted lock, and the next waiting (blocked) routine executing its first command. This experiment was performed with two instances of the same routine triggered back to back (i.e., all devices are conflicting devices). {We observe that  sync delay also rises slowly with system size---increasing  number of nodes by 15$\times$ from 50 to 750 causes $< 4 \times $ increase in sync delay (and client delay).  }

{One may be led to believe that client delay is higher than  sync delay. Yet, each has operations excluded in the other. Sync delay excludes triggering of  routine and associated \kgroup{} quorum. Client delay excludes lock release latency and associated %\kgroup{} 
quorums. }

\subsection{Load Distribution}

\begin{figure}[tb!]
    \centering
    \begin{subfigure}[b]{0.48\linewidth}
        \centering
        \includegraphics[width=\textwidth]{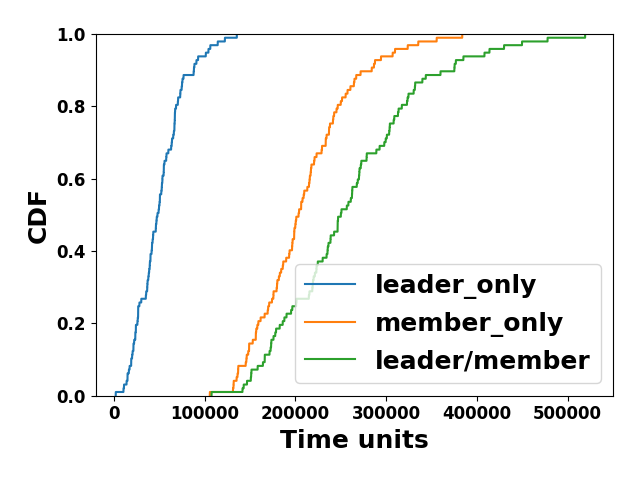}
        \vspace{-6pt}
        \caption{\textit{Temporal} load distribution with locality \kgroups{}}
        \label{subfig:temporal-load-balance-locality}
    \end{subfigure}
    \hfill
    \begin{subfigure}[b]{0.48\linewidth}
        \centering
        \includegraphics[width=\textwidth]{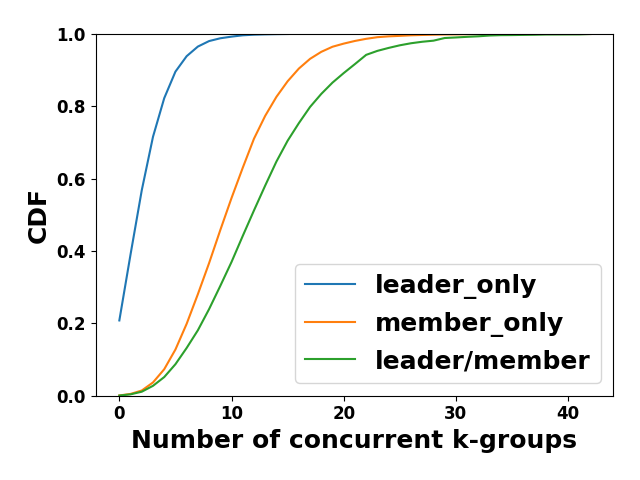}
        \vspace{-6pt}
        \caption{\textit{Spatial} load distribution with locality \kgroups{}}
        \label{subfig:spatial-load-balance-locality}
    \end{subfigure}
    \par\medskip
    \begin{subfigure}[b]{0.48\linewidth}
        \centering
        \includegraphics[width=\textwidth]{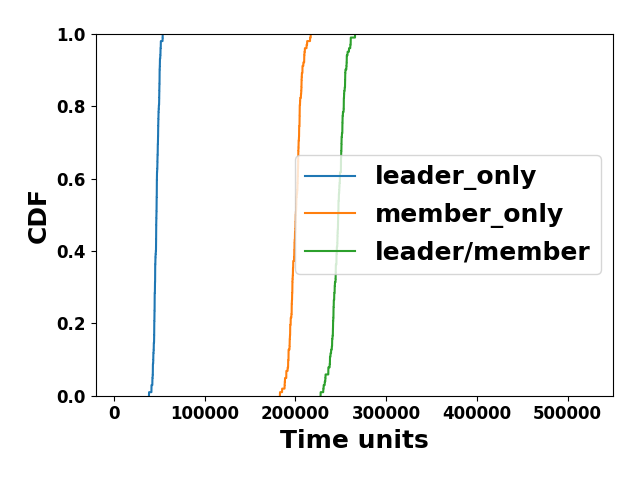}
        \vspace{-6pt}
        \caption{\textit{Temporal} load distribution with random \kgroups{}}
        \label{subfig:temporal-load-balance-random}
    \end{subfigure}
    \hfill
    \begin{subfigure}[b]{0.48\linewidth}
        \centering
        \includegraphics[width= \textwidth]{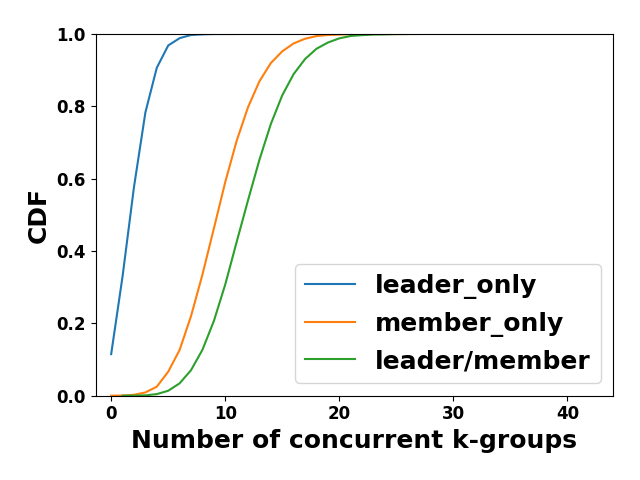}
        \vspace{-6pt}
        \caption{\textit{Spatial} load distribution with random \kgroups{}}
        \label{subfig:spatial-load-balance-random}
    \end{subfigure}
    \vspace{3pt}
    \caption{\small \it \sysname{}'s load distribution across time and space for the two member and leader selection algorithms (locality vs. random)}
    \label{fig:load-balance}
\end{figure}
\Figure~\ref{fig:load-balance} shows  the CDF of load distributions, in a grid topology, for the leader, for member, and for all (leader or member) nodes. { {\it Temporal load} is the amount of time that a smart device  serves in a particular role over the entire experiment duration.} 
{{\it Spatial load} is the number of concurrent k-groups that a smart device is present in.}

{The top two plots show \sysname{} with LSH and the bottom two plots show \sysname{} with random member selection. As expected, random member selection is the most load balanced across nodes and across time. Although LSH clusters responsibilities closer to target devices leading to load imbalance (top two plots), the median temporal load (left two plots) for member and leader/member are surprisingly comparable between random and LSH, showing that LSH balances member load almost as well as random. LSH member load has a higher tail due to clustering effects. The same clustering effects lead to repeated leader candidates and increase the median leader load in LSH (top left plot). 

}

\subsection{k-Group Operations}

\begin{figure}[tb]
    \centering
    \includegraphics[scale=0.45]{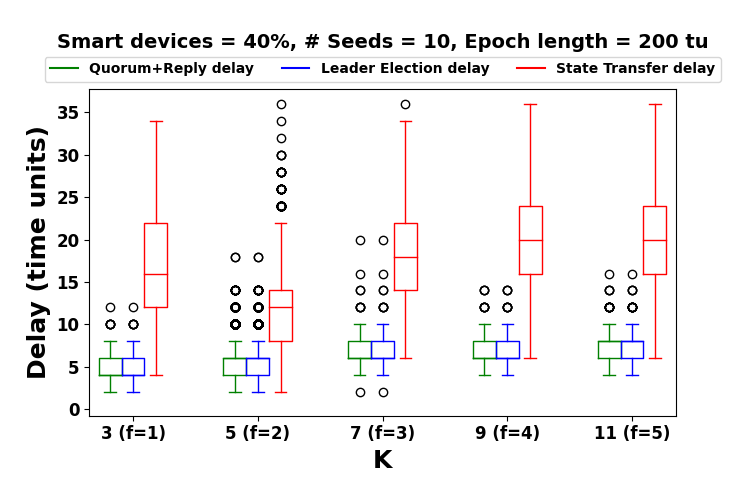}
    \vspace{-6pt}
    \caption{\small \it Delays of operations within a \kgroup{} for different values of $\K$
    }
    \label{fig:in_kgroup_ks}
\end{figure}

\begin{figure}[tb!]
    \centering
    \includegraphics[width=0.4 \textwidth]{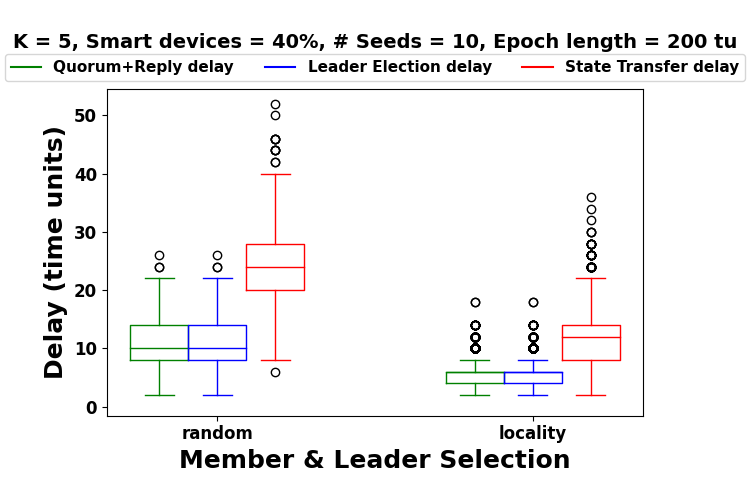}
    \vspace{-6pt}
    \caption{\small \it Delays of operations within a \kgroup{} for different policies for entity clustering, member selection, and leader election}
    \label{fig:in_kgroup_policies}
\end{figure}

\begin{figure}[tb!]
    \centering
    \includegraphics[scale=0.4]{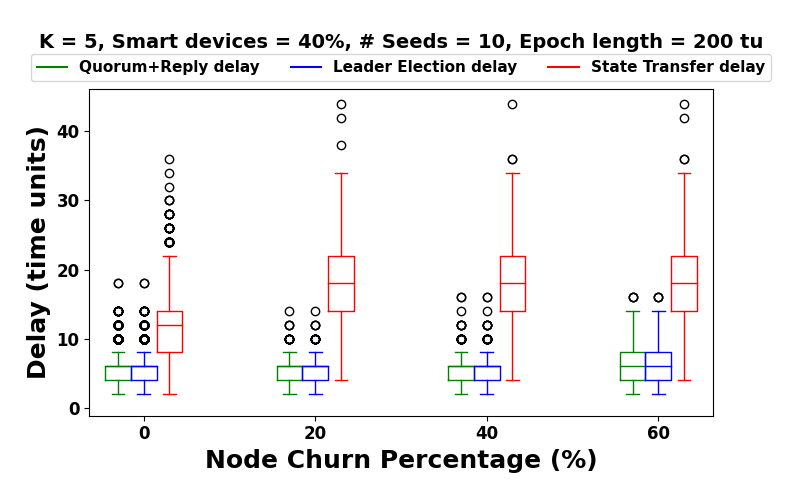}
    \vspace{-6pt}
    \caption{\small \it Delays of operations within a \kgroup{} for different percentages of node churn}
    \label{fig:in_kgroup_failure}
\end{figure}

We now measure the internal performance within \kgroups{}. We use the default parameters of Table~\ref{tab:def_param_vals}. \Figure~\ref{fig:in_kgroup_ks} shows that all \kgroup{} operations---election, quorum, and state transfer---scale  well as the size of the \kgroup{} is increased from 3 ($f=1$) to 11 ($f=5$). Quorums are scalable because requests and replies are sent in parallel. Leader election delays are scalable in the common case because of the best case of the Bully algorithm. State transfer delay is a function of the amount of state, but does not depend on $k$.

\Figure~\ref{fig:in_kgroup_policies} confirms our hypothesis that locality-aware
mechanisms for \kgroup{} entity selection, member selection,  and leader election
decrease the operational delays of a \kgroup{}.
The state transfer operation's delay is not decreased as much as the quorum or
leader election's delay since, apart from a quorum, the unicast to the old leader is not affected by LSH. 

\Figure~\ref{fig:in_kgroup_failure} shows  the k-group's operations are not 
affected by the \smart{}s' churn in the topology.
Only when the churn percentage affects about 60\% of \smart{}s 
do  \kgroup{} operations' get slower.
This is because a \kgroup's operation only requires a majority of its members. 
Upon the failure of $\f$ members in a \kgroup{}, the leader ensures that  failed
members are replaced so that future failures of up to $\f$ \smart{}s do not affect it either.

\section{Deployment Evaluation}
\label{sec:deployment}

We implemented \sysname{} in the Raspberry Pi (RPi) 4  environment. Our implementation contains {12.5K lines of Java code}. Our deployments are on  RPi 4 model B,
possessing 2GB of LPDDR4 RAM and a Broadcom BCM2711 quad-core CPU of 1.5GHz Cortex-A72 cores. While Section~\ref{section:exp_eval} showed larger-scale simulation results, our goal with RPi's is to  benchmark the behavior of \sysname's core mechanism, i.e., k-groups, on a small number of real devices. 

We deployed 11 RPi 4 devices in our lab, forming different mesh topologies. To attenuate RPi's strong signal and create the mesh topology, we consistently wrap each device in aluminum foil to reduce transmission power to  the lowest setting, 15dBm~\cite{Safehome}.  While \sysname{} works
 with any ad-hoc routing protocol, for concreteness
we use OLSR routing~\cite{olsrd} due to its standardization in Pi 4s.

We configure \sysname{} to use underneath it the  open-source implementation of the  Medley~\cite{medley}  membership protocol.
\sysname{} is configured to use \kgroups{} with $k=3$, and monitors devices every second. 
To avoid cold start biases, we start measurements 10 s after \sysname{} is initialized. Each data point involves at least 5 trial runs with different seeds.
Epochs change every 60 s. Different routines may be triggered simultaneously. The same routine may be triggered multiple times, but never within less than 10 s of its own previous invocation.  
We use SLA locking (\Section{}\ref{subsubsec:sla}).

\begin{figure}[tb!]
    \centering
    \begin{subfigure}[t!]{0.2\textwidth}
        \centering
        \includegraphics[width=\textwidth]{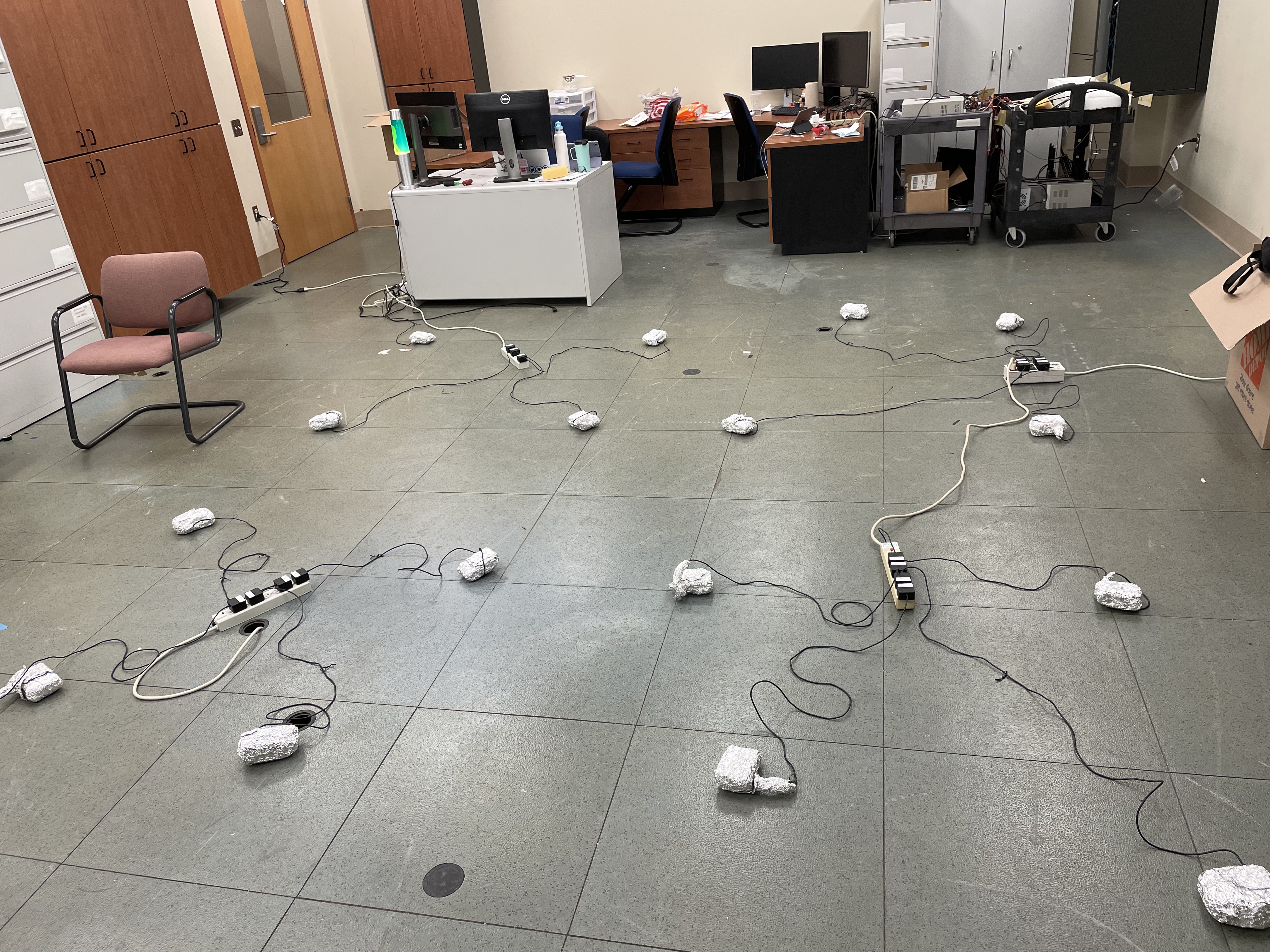}
        \caption{Grid Topology}
        \label{subfig:grid_topo}
    \end{subfigure}
    \begin{subfigure}[t!]{0.2\textwidth}
        \centering
        \includegraphics[width=\textwidth,angle=-90,origin=c]{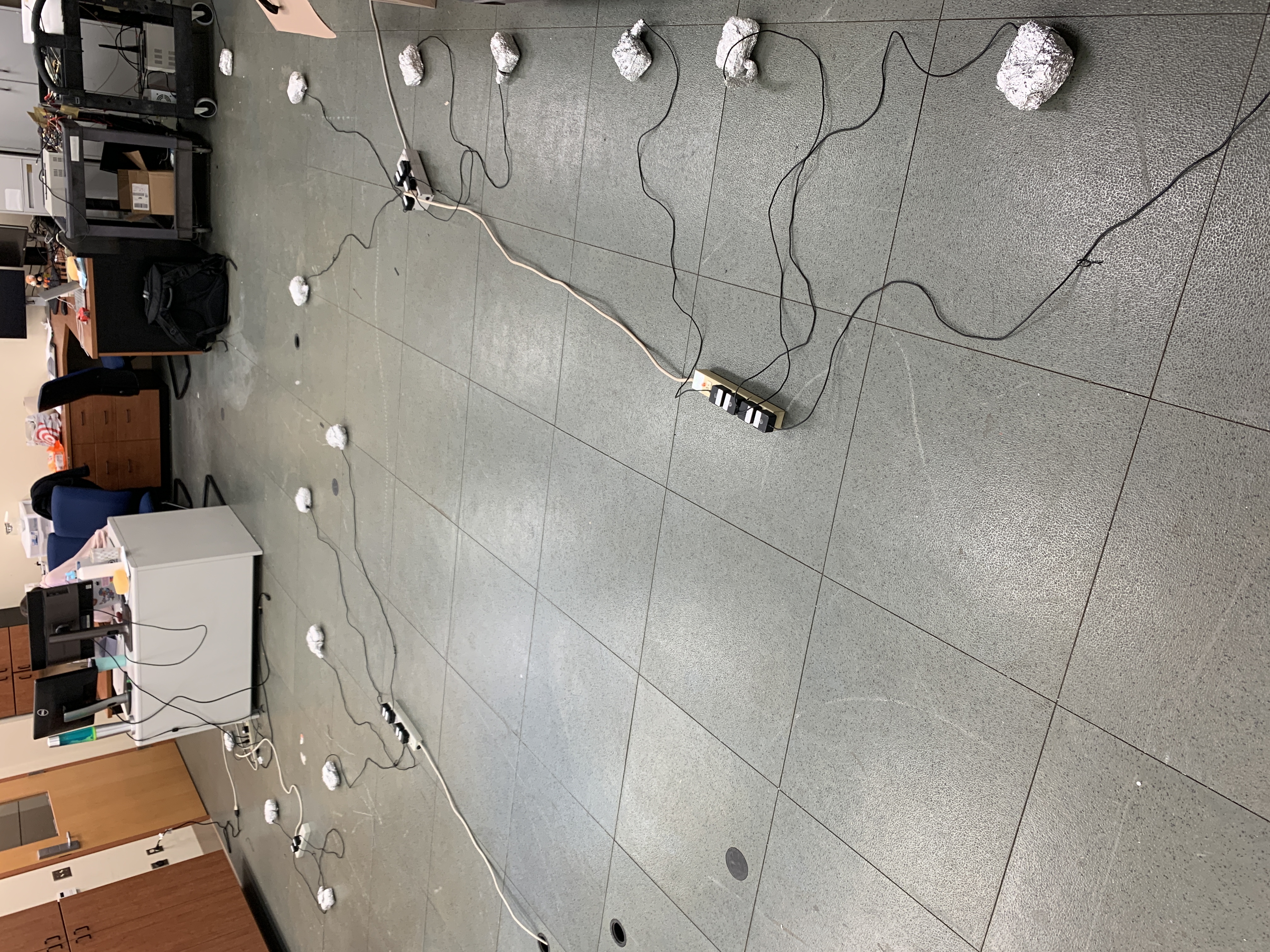}
        \caption{Line Topology}
        \label{subfig:line_topo}
    \end{subfigure}
    \vspace{-6pt}
    \caption{\small \it Raspberry Pi Topologies: Lab Deployment.}
    \label{fig:rpi_topos}
\end{figure}

We implement two Pi topologies shown in \Figure~\ref{fig:rpi_topos}: (i) a 4 $\times$ 4 2D {\it Grid}-like topology and (ii) an L-shaped {\it Line} (modeling the outline of one floor of our office building). Each covers an area of  4 meters $\times$ 4 meters. 
The Line topology's diameter is higher than Grid's.

\begin{figure}[tb!]
    \centering
    \includegraphics[width=0.5 \textwidth]{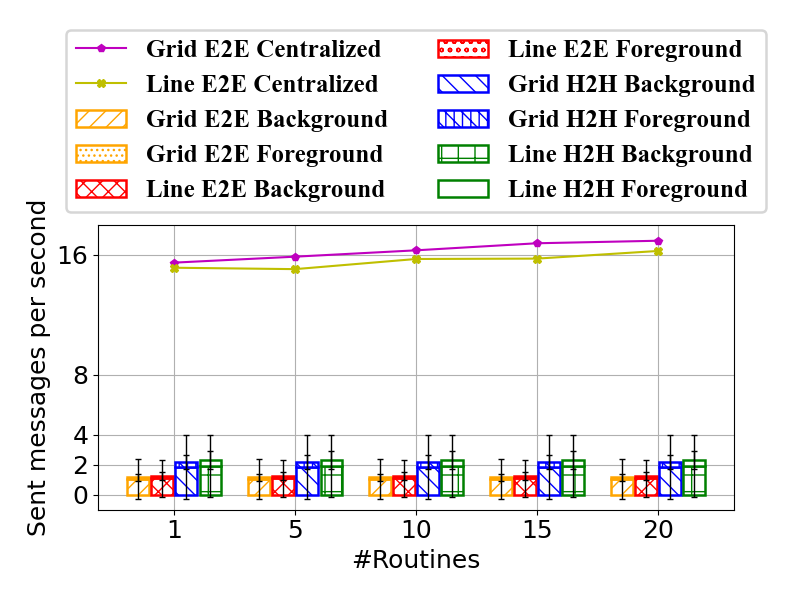}
    \vspace{-25pt}
    \caption{\small \it Bandwidth: Decentralized \sysname{} vs. Centralized.}
    \label{fig:bw_centralized_skytali}
\end{figure}

\noindent{\bf Bandwidth Usage---Decentralized vs. Centralized: } 
\Figure{}~\ref{fig:bw_centralized_skytali} shows decentralized \sysname's bandwidth per node, both  end to end (E2E) messages, as well as ad-hoc routed (hop to hop/ H2H) messages. For comparison we also instantiated \sysname{} with $k=1$ and one \kgroup, i.e., a single centralized device  responsible for all routines' monitoring and management---this is akin to using a central home hub (\Section{}~\ref{section:intro}). We observe that: a) Decentralized \sysname{} reduces bandwidth by an order of magnitude (over 10 $\times$) compared to the centralized approach. b) Both end to end bandwidth and hop bandwidth are scalable, i.e., remain flat, as the number of routines grows. c) Bandwidth of foreground operations (e.g.,  election, k-group migration, etc.) remain small compared to unavoidable background bandwidth (e.g., device monitoring).

\begin{figure}[tb!]
    \centering
    \includegraphics[width=0.5 \textwidth]{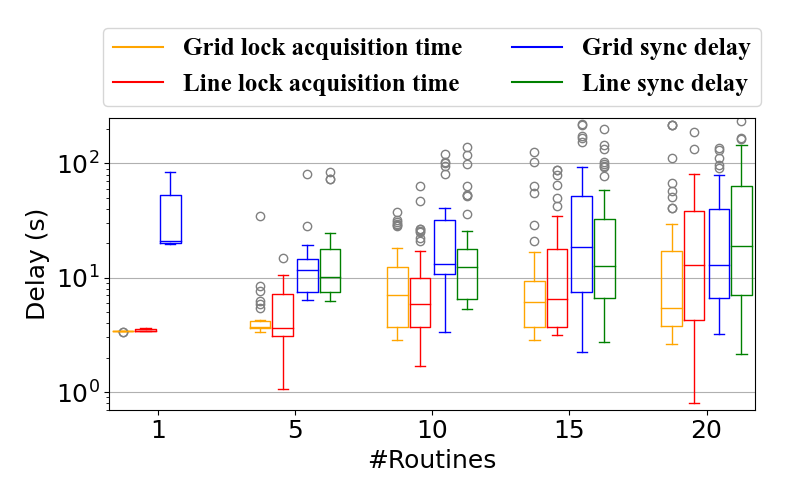}
    \vspace{-25pt}
    \caption{\small \it Lock Acquisition Time and Sync Delay. }
    \label{fig:ala_sync_mul_rtns}
\end{figure}

\noindent{\bf Wait Times to Start a Routine: }
\Figure{}~\ref{fig:ala_sync_mul_rtns} shows two metrics  for any given routine. First, the Lock Acquisition time---the average time to acquire a device lock after it becomes available---varies
between 3.41 s - 13.02 s.
As routine count increases,  median
latencies  remain  stable for Grid, and rise slightly for Line. 
Second, the Sync delay---the time between the routine's last device becoming available and the routine starting---stays  stable with increasing routine count, with  median  between 10.23 s - 20.833 s.

\noindent{\bf Inter-\kgroup{} Operation Latencies: }
\Table{}~\ref{tab:kgroup_latencies} shows that \sysname's common operations are fast and take a median of 1.08 s for quorum, 1.17 s for election, and 2.07 s for state transfer during epoch change.

\begin{table}[]
{\small
    \centering
    \begin{tabular}{c||c|c|c}
        \kgroup{} delay & median & mean & P90 \\
        \hline
        Quorum reply & 1.0805 & 1.4021 & 1.4962 \\
        Leader election & 1.1675 & 1.4913 & 1.9249 \\
        State transfer & 2.0750 & 2.8468 & 3.1390
        \vspace{5pt}
    \end{tabular}
    }
     \caption{\it \small \sysname's \kgroup{} Latencies  (seconds). }
    \label{tab:kgroup_latencies}
\end{table}
\vspace{-5pt}

\section{Related Work}
\label{section:related_work}

\noindent \textbf{Consistent Hashing \& Committees:  }
Peer-to-peer (p2p) networked systems such as Chord~\cite{chord}, Dynamo~\cite{dynamo},
Cassandra~\cite{cassandra} employ consistent hashing~\cite{consistent}
to find the node storing a key-value pair.
\sysname{}'s \kgroup{} mechanism is inspired by consistent subgroups used by the election protocol in \cite{large_election}, but that mechanism is neither dynamic, nor intended for IoT settings, nor has locality.

\noindent \textbf{Consensus: }
A variety of agreement protocols for wireless environment have been proposed, such as 
Wireless Paxos~\cite{wirelesspaxos}, and  WSN Byzantine consensus~\cite{wchain}.
Unlike these consensus algorithms which require many message exchange rounds, 
\sysname{}'s \kgroups{} are efficient and fast, and work quickly in the common case,  with just one communication round for a quorum.

\noindent \textbf{IoT Application Management: }
IoT routine and application management system have  been explored widely, but with a centralized control plane, e.g.,  HomeOS~\cite{HomeOS},  Beam~\cite{Beam}, 
DepSys~\cite{DepSys},  eHome~\cite{DepManagement}, and 
SafeHome~\cite{Safehome}.
Rivulet~\cite{rivulet} is distributed and uses \smart{}s to exchange messages and backup state to shadow nodes. However, unlike \sysname, Rivulet focuses on fault-tolerance and omits the routine management.
Other works on IoT fault-tolerance~\cite{Wukong,Cefiot,iotrepair}, also do not manage routine execution. 
To the best of our knowledge, \sysname{} is the first fully-decentralized IoT routine management system.

\section{Summary}
\label{section:conclusion}

We have presented the \sysname{} system that manages routines in a smart space (building, home, or campus) in a fully-decentralized way, without relying on any central components. Analysis of \sysname's techniques show that it achieves  continuity (inheritance), safety, liveness,  and progress. Our implementation and simulation of \sysname{} under various topologies show that it scales well with increasing number of devices into the hundreds, and with larger values of $f$ (number of failures). \sysname{} also balances load effectively, both over smart devices and over time. \sysname{} opens the door for realizing self-managing versions of today's  smart space systems. Future directions include building central-distributed hybrid management systems, accounting for heterogeneity of smart devices, and energy considerations.

% \newpage
\bibliographystyle{ACM-Reference-Format}
\bibliography{paper}


\begin{thebibliography}{62}

%%% the Web address.

\ifx \showCODEN    \undefined \def \showCODEN     #1{\unskip}     \fi
\ifx \showDOI      \undefined \def \showDOI       #1{#1}\fi
\ifx \showISBNx    \undefined \def \showISBNx     #1{\unskip}     \fi
\ifx \showISBNxiii \undefined \def \showISBNxiii  #1{\unskip}     \fi
\ifx \showISSN     \undefined \def \showISSN      #1{\unskip}     \fi
\ifx \showLCCN     \undefined \def \showLCCN      #1{\unskip}     \fi
\ifx \shownote     \undefined \def \shownote      #1{#1}          \fi
\ifx \showarticletitle \undefined \def \showarticletitle #1{#1}   \fi
\ifx \showURL      \undefined \def \showURL       {\relax}        \fi
% invisible to TeX
\providecommand\bibfield[2]{#2}
\providecommand\bibinfo[2]{#2}
\providecommand\natexlab[1]{#1}
\providecommand\showeprint[2][]{arXiv:#2}

\bibitem[rpi(2019)]%
        {rpi-slow-speed1}
 \bibinfo{year}{2019}\natexlab{}.
\newblock \bibinfo{title}{Slow upload on new Raspberry Pi 4 2GB}.
\newblock
\newblock
\urldef\tempurl%
\url{https://forums.raspberrypi.com/viewtopic.php?t=244832}
\showURL{%
\tempurl}


\bibitem[Aarti and Vineet(2022)]%
        {iotbattlemarket}
\bibfield{author}{\bibinfo{person}{G. Aarti} {and} \bibinfo{person}{K.
  Vineet}.} \bibinfo{year}{2022}\natexlab{}.
\newblock \bibinfo{title}{IoT in Aerospace \& Defense Market by Component
  (Hardware, Software, Services), by Deployment Mode (On-Premise, Cloud), by
  Connectivity Technology (Cellular, Wi-Fi, Satellite Communication, Radio
  Frequency), by Application (Fleet Management, Inventory Management, Equipment
  Maintenance, Security, Others): Global Opportunity Analysis and Industry
  Forecast, 2020-2030}.
\newblock
  \bibinfo{howpublished}{https://www.alliedmarketresearch.com/internet-of-things-in-aerospace-and-defense-market}.
\newblock


\bibitem[Ahsan et~al\mbox{.}(2021)]%
        {Safehome}
\bibfield{author}{\bibinfo{person}{S.~B. Ahsan}, \bibinfo{person}{R. Yang},
  \bibinfo{person}{S.~A. Noghabi}, {and} \bibinfo{person}{I. Gupta}.}
  \bibinfo{year}{2021}\natexlab{}.
\newblock \showarticletitle{Home, {SafeHome}: smart home reliability with
  visibility and atomicity}. In \bibinfo{booktitle}{\emph{EuroSys}}.
\newblock


\bibitem[Akyildiz and Wang(2005)]%
        {akyildiz2005survey}
\bibfield{author}{\bibinfo{person}{I.~F. Akyildiz} {and} \bibinfo{person}{X.
  Wang}.} \bibinfo{year}{2005}\natexlab{}.
\newblock \showarticletitle{A survey on wireless mesh networks}.
\newblock \bibinfo{journal}{\emph{IEEE Commun. Mag.}} \bibinfo{volume}{43},
  \bibinfo{number}{9} (\bibinfo{year}{2005}).
\newblock


\bibitem[Amazon Alexa(2014)]%
        {Alexa}
Amazon Alexa \bibinfo{year}{2014}\natexlab{}.
\newblock \bibinfo{title}{{A}mazon {A}lexa}.
\newblock \bibinfo{howpublished}{\url{https://developer.amazon.com/alexa}}.
\newblock


\bibitem[Apecechea(2019)]%
        {cryptosort}
\bibfield{author}{\bibinfo{person}{G.~I. Apecechea}.}
  \bibinfo{year}{2019}\natexlab{}.
\newblock \bibinfo{title}{Cryptographic Sortition in Blockchains: the
  importance of VRFs}.
\newblock
\newblock
\urldef\tempurl%
\url{https://medium.com/witnet/cryptographic-sortition-in-blockchains-the-importance-of-vrfs-ad5c20a4e018}
\showURL{%
\tempurl}


\bibitem[Ardekani et~al\mbox{.}(2017)]%
        {rivulet}
\bibfield{author}{\bibinfo{person}{M. Ardekani}, \bibinfo{person}{R. Singh},
  \bibinfo{person}{N. Agrawal}, \bibinfo{person}{D. Terry}, {and}
  \bibinfo{person}{R. Suminto}.} \bibinfo{year}{2017}\natexlab{}.
\newblock \showarticletitle{Rivulet: A Fault-Tolerant Platform for Smart-Home
  Applications}. In \bibinfo{booktitle}{\emph{Middleware}}.
\newblock


\bibitem[{AWS}(2022)]%
        {AWSIoT}
\bibfield{author}{\bibinfo{person}{{AWS}}.} \bibinfo{year}{2022}\natexlab{}.
\newblock \bibinfo{title}{AWS IoT Core additional metering details}.
\newblock
\newblock
\urldef\tempurl%
\url{https://aws.amazon.com/iot-core/pricing/additional-details/}
\showURL{%
\tempurl}


\bibitem[Chandra and Collis(2021)]%
        {chandra21}
\bibfield{author}{\bibinfo{person}{R. Chandra} {and} \bibinfo{person}{S.
  Collis}.} \bibinfo{year}{2021}\natexlab{}.
\newblock \showarticletitle{Digital Agriculture for Small-Scale Producers:
  Challenges and Opportunities}.
\newblock \bibinfo{journal}{\emph{CACM}} \bibinfo{volume}{64},
  \bibinfo{number}{12} (\bibinfo{year}{2021}).
\newblock


\bibitem[Datar et~al\mbox{.}(2004)]%
        {datar_locality-sensitive_2004}
\bibfield{author}{\bibinfo{person}{M. Datar}, \bibinfo{person}{N. Immorlica},
  \bibinfo{person}{P. Indyk}, {and} \bibinfo{person}{V.~S. Mirrokni}.}
  \bibinfo{year}{2004}\natexlab{}.
\newblock \showarticletitle{Locality-sensitive hashing scheme based on p-stable
  distributions}. In \bibinfo{booktitle}{\emph{SoCG}}.
\newblock


\bibitem[DeCandia et~al\mbox{.}(2007)]%
        {dynamo}
\bibfield{author}{\bibinfo{person}{G. DeCandia}, \bibinfo{person}{D. Hastorun},
  \bibinfo{person}{M. Jampani}, \bibinfo{person}{G. Kakulapati},
  \bibinfo{person}{A. Lakshman}, \bibinfo{person}{A. Pilchin},
  \bibinfo{person}{S. Sivasubramanian}, \bibinfo{person}{P. Vosshall}, {and}
  \bibinfo{person}{W. Vogels}.} \bibinfo{year}{2007}\natexlab{}.
\newblock \showarticletitle{Dynamo: Amazon's Highly Available Key-Value Store}.
  In \bibinfo{booktitle}{\emph{SOSP}}.
\newblock


\bibitem[Dijkstra(1959)]%
        {dijkstra}
\bibfield{author}{\bibinfo{person}{E.~W. et~al. Dijkstra}.}
  \bibinfo{year}{1959}\natexlab{}.
\newblock \showarticletitle{A note on two problems in connexion with graphs}.
\newblock \bibinfo{journal}{\emph{Numerische mathematik}} \bibinfo{volume}{1},
  \bibinfo{number}{1} (\bibinfo{year}{1959}).
\newblock


\bibitem[Dixon et~al\mbox{.}(2012)]%
        {HomeOS}
\bibfield{author}{\bibinfo{person}{C. Dixon}, \bibinfo{person}{R. Mahajan},
  \bibinfo{person}{S. Agarwal}, \bibinfo{person}{A.~J. Brush},
  \bibinfo{person}{B. Lee}, \bibinfo{person}{S. Saroiu}, {and}
  \bibinfo{person}{P. Bahl}.} \bibinfo{year}{2012}\natexlab{}.
\newblock \showarticletitle{An operating system for the home}. In
  \bibinfo{booktitle}{\emph{NSDI}}.
\newblock


\bibitem[Fatima et~al\mbox{.}(2022)]%
        {fatima2022production}
\bibfield{author}{\bibinfo{person}{Z. Fatima}, \bibinfo{person}{M.~H. Tanveer},
  \bibinfo{person}{S. Zardari}, \bibinfo{person}{L.~F. Naz},
  \bibinfo{person}{H. Khadim}, \bibinfo{person}{N. Ahmed}, {and}
  \bibinfo{person}{M. Tahir}.} \bibinfo{year}{2022}\natexlab{}.
\newblock \showarticletitle{Production plant and warehouse automation with IoT
  and industry 5.0}.
\newblock \bibinfo{journal}{\emph{Applied Sciences}} \bibinfo{volume}{12},
  \bibinfo{number}{4} (\bibinfo{year}{2022}).
\newblock


\bibitem[Feng et~al\mbox{.}(2020)]%
        {feng2020robustness}
\bibfield{author}{\bibinfo{person}{Y. Feng}, \bibinfo{person}{M. Li},
  \bibinfo{person}{C. Zeng}, {and} \bibinfo{person}{H. Liu}.}
  \bibinfo{year}{2020}\natexlab{}.
\newblock \showarticletitle{Robustness of internet of battlefield things
  (iobt): A directed network perspective}.
\newblock \bibinfo{journal}{\emph{Entropy}} \bibinfo{volume}{22},
  \bibinfo{number}{10} (\bibinfo{year}{2020}).
\newblock


\bibitem[Fox et~al\mbox{.}(2019)]%
        {fox2019live}
\bibfield{author}{\bibinfo{person}{M.~A. Fox}, \bibinfo{person}{J.~L. Breese},
  {and} \bibinfo{person}{G. Vaidyanathan}.} \bibinfo{year}{2019}\natexlab{}.
\newblock \bibinfo{booktitle}{\emph{Live Music Performances and the Internet of
  Things}}.
\newblock


\bibitem[Future(2022)]%
        {iotstadiummarket}
\bibfield{author}{\bibinfo{person}{Market~Research Future}.}
  \bibinfo{year}{2022}\natexlab{}.
\newblock \bibinfo{title}{Smart Stadium Market Estimated to Hit 24.3 Billion
  with a CAGR of 23\% during 2021 - 2030}.
\newblock
  \bibinfo{howpublished}{https://www.globenewswire.com/news-release/2022/06/01/2453900/0/en/Smart-Stadium-Market-Estimated-to-Hit-24-3-Billion-with-a-CAGR-of-23-during-2021-2030-Report-by-Market-Research-Future-MRFR.html}.
\newblock


\bibitem[Gan and Lee(2018)]%
        {gan2018}
\bibfield{author}{\bibinfo{person}{H. Gan} {and} \bibinfo{person}{W.S. Lee}.}
  \bibinfo{year}{2018}\natexlab{}.
\newblock \showarticletitle{Development of a Navigation System for a Smart
  Farm}.
\newblock \bibinfo{journal}{\emph{IFAC-PapersOnLine}} \bibinfo{volume}{51},
  \bibinfo{number}{17} (\bibinfo{year}{2018}).
\newblock


\bibitem[Garcia-Molina(1982)]%
        {Bully}
\bibfield{author}{\bibinfo{person}{H. Garcia-Molina}.}
  \bibinfo{year}{1982}\natexlab{}.
\newblock \showarticletitle{Elections in a Distributed Computing System}.
\newblock \bibinfo{journal}{\emph{IEEE TOC}} \bibinfo{volume}{C-31},
  \bibinfo{number}{1} (\bibinfo{year}{1982}).
\newblock


\bibitem[Google Home(2016)]%
        {GoogleHome}
Google Home \bibinfo{year}{2016}\natexlab{}.
\newblock \bibinfo{title}{{G}oogle {H}ome}.
\newblock
  \bibinfo{howpublished}{\url{https://store.google.com/us/product/google_home}}.
\newblock


\bibitem[Grothaus(2019)]%
        {gcloud-down1}
\bibfield{author}{\bibinfo{person}{M. Grothaus}.}
  \bibinfo{year}{2019}\natexlab{}.
\newblock \bibinfo{title}{That major Google outage meant some Nest users
  couldn’t unlock doors or use the AC}.
\newblock
\newblock
\urldef\tempurl%
\url{https://www.fastcompany.com/90358396/that-major-google-outage-meant-some-nest-users-couldnt-unlock-doors-or-use-the-ac}
\showURL{%
\tempurl}


\bibitem[Guo et~al\mbox{.}(2022)]%
        {sayian}
\bibfield{author}{\bibinfo{person}{X. Guo}, \bibinfo{person}{Lo. Shangguan},
  \bibinfo{person}{Y. He}, \bibinfo{person}{N. Jing}, \bibinfo{person}{J.
  Zhang}, \bibinfo{person}{H. Jiang}, {and} \bibinfo{person}{Y. Liu}.}
  \bibinfo{year}{2022}\natexlab{}.
\newblock \showarticletitle{Saiyan: Design and Implementation of a Low-power
  Demodulator for {LoRa} Backscatter Systems}. In
  \bibinfo{booktitle}{\emph{NSDI}}.
\newblock


\bibitem[Gupta et~al\mbox{.}(2000)]%
        {large_election}
\bibfield{author}{\bibinfo{person}{I. Gupta}, \bibinfo{person}{R. Van~Renesse},
  {and} \bibinfo{person}{K.~P Birman}.} \bibinfo{year}{2000}\natexlab{}.
\newblock \showarticletitle{A probabilistically correct leader election
  protocol for large groups}. In \bibinfo{booktitle}{\emph{DISC}}.
\newblock


\bibitem[Hunt et~al\mbox{.}(2010)]%
        {zookeeper}
\bibfield{author}{\bibinfo{person}{P. Hunt}, \bibinfo{person}{M. Konar},
  \bibinfo{person}{F.~P. Junqueira}, {and} \bibinfo{person}{B. Reed}.}
  \bibinfo{year}{2010}\natexlab{}.
\newblock \showarticletitle{ZooKeeper: Wait-free Coordination for
  Internet-scale Systems.}. In \bibinfo{booktitle}{\emph{ATC}}.
\newblock


\bibitem[{ISO Central Secretary}(2016)]%
        {iso_central_secretary_systems_2016}
\bibfield{author}{\bibinfo{person}{{ISO Central Secretary}}.}
  \bibinfo{year}{2016}\natexlab{}.
\newblock \bibinfo{booktitle}{\emph{Systems and software engineering --
  {Lifecycle} profiles for {Very} {Small} {Entities} ({VSEs}) -- {Part} 1:
  {Overview}}}.
\newblock \bibinfo{type}{Standard} ISO/IEC TR 29110-1:2016.
  \bibinfo{institution}{International Organization for Standardization},
  \bibinfo{address}{Geneva, CH}.
\newblock
\urldef\tempurl%
\url{https://www.iso.org/standard/62711.html}
\showURL{%
\tempurl}


\bibitem[J.(2021)]%
        {IoTstats}
\bibfield{author}{\bibinfo{person}{Bojan J.}} \bibinfo{year}{2021}\natexlab{}.
\newblock \bibinfo{title}{{Internet of Things} statistics for 2021 – Taking
  Things Apart}.
\newblock
  \bibinfo{howpublished}{\url{https://dataprot.net/statistics/iot-statistics/}}.
\newblock


\bibitem[Javed et~al\mbox{.}(2018)]%
        {Cefiot}
\bibfield{author}{\bibinfo{person}{A. Javed}, \bibinfo{person}{K. Heljanko},
  \bibinfo{person}{A. Buda}, {and} \bibinfo{person}{K. Fr{\"a}mling}.}
  \bibinfo{year}{2018}\natexlab{}.
\newblock \showarticletitle{{CEFIoT}: A fault-tolerant {IoT} architecture for
  edge and cloud}. In \bibinfo{booktitle}{\emph{WF-IoT}}.
\newblock


\bibitem[Johnson and Maltz(1996)]%
        {dsr}
\bibfield{author}{\bibinfo{person}{D.~B. Johnson} {and} \bibinfo{person}{D.~A.
  Maltz}.} \bibinfo{year}{1996}\natexlab{}.
\newblock \showarticletitle{Dynamic source routing in ad hoc wireless
  networks}.
\newblock In \bibinfo{booktitle}{\emph{Mobile computing}}.
\newblock


\bibitem[K. et~al\mbox{.}(2021)]%
        {katanbaf21}
\bibfield{author}{\bibinfo{person}{Mohamad K.}, \bibinfo{person}{Anthony W.},
  {and} \bibinfo{person}{Vamsi T.}} \bibinfo{year}{2021}\natexlab{}.
\newblock \showarticletitle{Simplifying Backscatter Deployment: {Full-Duplex}
  {LoRa} Backscatter}. In \bibinfo{booktitle}{\emph{NSDI}}.
\newblock


\bibitem[Karger et~al\mbox{.}(1997)]%
        {consistent}
\bibfield{author}{\bibinfo{person}{David Karger}, \bibinfo{person}{Eric
  Lehman}, \bibinfo{person}{Tom Leighton}, \bibinfo{person}{Rina Panigrahy},
  \bibinfo{person}{Matthew Levine}, {and} \bibinfo{person}{Daniel Lewin}.}
  \bibinfo{year}{1997}\natexlab{}.
\newblock \showarticletitle{Consistent hashing and random trees: Distributed
  caching protocols for relieving hot spots on the world wide web}. In
  \bibinfo{booktitle}{\emph{STOC}}. \bibinfo{pages}{654--663}.
\newblock


\bibitem[Krishna and Narasimha~Murty(1999)]%
        {KMeans}
\bibfield{author}{\bibinfo{person}{K. Krishna} {and} \bibinfo{person}{M.
  Narasimha~Murty}.} \bibinfo{year}{1999}\natexlab{}.
\newblock \showarticletitle{Genetic K-means algorithm}.
\newblock \bibinfo{journal}{\emph{IEEE Transactions on Systems, Man, and
  Cybernetics, Part B (Cybernetics)}} \bibinfo{volume}{29}, \bibinfo{number}{3}
  (\bibinfo{year}{1999}), \bibinfo{pages}{433--439}.
\newblock
\urldef\tempurl%
\url{https://doi.org/10.1109/3477.764879}
\showDOI{\tempurl}


\bibitem[Lakshman and Malik(2010)]%
        {cassandra}
\bibfield{author}{\bibinfo{person}{A. Lakshman} {and} \bibinfo{person}{P.
  Malik}.} \bibinfo{year}{2010}\natexlab{}.
\newblock \showarticletitle{Cassandra: a decentralized structured storage
  system}.
\newblock \bibinfo{journal}{\emph{IGOPS}} \bibinfo{volume}{44},
  \bibinfo{number}{2} (\bibinfo{year}{2010}).
\newblock


\bibitem[Lampropoulos et~al\mbox{.}(2019)]%
        {lampropoulos2019internet}
\bibfield{author}{\bibinfo{person}{G. Lampropoulos}, \bibinfo{person}{Siakas
  K.}, {and} \bibinfo{person}{T. Anastasiadis}.}
  \bibinfo{year}{2019}\natexlab{}.
\newblock \showarticletitle{Internet of things in the context of industry 4.0:
  An overview}.
\newblock \bibinfo{journal}{\emph{IJEK}} (\bibinfo{year}{2019}).
\newblock


\bibitem[Li et~al\mbox{.}(2022)]%
        {curvinglora}
\bibfield{author}{\bibinfo{person}{Chenning Li}, \bibinfo{person}{Xiuzhen Guo},
  \bibinfo{person}{Longfei Shangguan}, \bibinfo{person}{Zhichao Cao}, {and}
  \bibinfo{person}{Kyle Jamieson}.} \bibinfo{year}{2022}\natexlab{}.
\newblock \showarticletitle{{CurvingLoRa} to Boost {LoRa} Network Throughput
  via Concurrent Transmission}. In \bibinfo{booktitle}{\emph{NSDI}}.
\newblock


\bibitem[Lin et~al\mbox{.}(2016)]%
        {lin2016human}
\bibfield{author}{\bibinfo{person}{K. Lin}, \bibinfo{person}{W. Wang},
  \bibinfo{person}{M. Bi, Y.and~Qiu}, {and} \bibinfo{person}{M.~M. Hassan}.}
  \bibinfo{year}{2016}\natexlab{}.
\newblock \showarticletitle{Human localization based on inertial sensors and
  fingerprints in the Industrial Internet of Things}.
\newblock \bibinfo{journal}{\emph{Elsevier COMNET}}  \bibinfo{volume}{101}
  (\bibinfo{year}{2016}).
\newblock


\bibitem[Liu et~al\mbox{.}(2022)]%
        {IoBT}
\bibfield{author}{\bibinfo{person}{D. Liu}, \bibinfo{person}{T.~F. Abdelzaher},
  \bibinfo{person}{T. Wang}, \bibinfo{person}{Y. Hu}, \bibinfo{person}{J. Li},
  \bibinfo{person}{S. Liu}, \bibinfo{person}{M. Caesar}, \bibinfo{person}{D.
  Kalasapura}, \bibinfo{person}{J. Bhattacharyya}, \bibinfo{person}{N. Srour},
  \bibinfo{person}{J. Kim}, \bibinfo{person}{G. Wang}, \bibinfo{person}{G.
  Kimberly}, {and} \bibinfo{person}{S. Yao}.} \bibinfo{year}{2022}\natexlab{}.
\newblock \showarticletitle{IoBT-OS: Optimizing the Sensing-to-Decision Loop
  for the Internet of Battlefield Things}. In
  \bibinfo{booktitle}{\emph{ICCCN}}.
\newblock


\bibitem[McNally(2022)]%
        {google-nest-speed}
\bibfield{author}{\bibinfo{person}{C. McNally}.}
  \bibinfo{year}{2022}\natexlab{}.
\newblock \bibinfo{title}{Google Nest Wi-Fi Review 2022}.
\newblock
\newblock
\urldef\tempurl%
\url{https://www.reviews.org/internet-service/google-nest-wifi-review/}
\showURL{%
\tempurl}


\bibitem[Memoori(2022)]%
        {Memoori}
\bibfield{author}{\bibinfo{person}{Memoori}.} \bibinfo{year}{2022}\natexlab{}.
\newblock \bibinfo{title}{{The Global Market for the Internet of Things in
  Smart Commercial Buildings}}.
\newblock
  \bibinfo{howpublished}{\url{https://memoori.com/portfolio/the-internet-of-things-in-smart-commercial-buildings-2022-to-2027/
  }}.
\newblock


\bibitem[Micali et~al\mbox{.}(1999)]%
        {micali1999verifiable}
\bibfield{author}{\bibinfo{person}{S. Micali}, \bibinfo{person}{M. Rabin},
  {and} \bibinfo{person}{S. Vadhan}.} \bibinfo{year}{1999}\natexlab{}.
\newblock \showarticletitle{Verifiable random functions}. In
  \bibinfo{booktitle}{\emph{FOCS}}.
\newblock


\bibitem[Moore et~al\mbox{.}(2020)]%
        {moore2020iot}
\bibfield{author}{\bibinfo{person}{S.~J. Moore}, \bibinfo{person}{C.~D.
  Nugent}, \bibinfo{person}{S. Zhang}, {and} \bibinfo{person}{I. Cleland}.}
  \bibinfo{year}{2020}\natexlab{}.
\newblock \showarticletitle{IoT reliability: a review leading to 5 key research
  directions}.
\newblock \bibinfo{journal}{\emph{CCF TPCI}}  \bibinfo{volume}{2}
  (\bibinfo{year}{2020}).
\newblock


\bibitem[Munir and Stankovic(2014)]%
        {DepSys}
\bibfield{author}{\bibinfo{person}{S. Munir} {and} \bibinfo{person}{J.~A.
  Stankovic}.} \bibinfo{year}{2014}\natexlab{}.
\newblock \showarticletitle{Dep{s}ys: Dependency aware integration of
  cyber-physical systems for smart homes}. In
  \bibinfo{booktitle}{\emph{ICCPS}}.
\newblock


\bibitem[Norris et~al\mbox{.}(2020)]%
        {iotrepair}
\bibfield{author}{\bibinfo{person}{M. Norris}, \bibinfo{person}{B. Celik},
  \bibinfo{person}{P. Venkatesh}, \bibinfo{person}{S. Zhao},
  \bibinfo{person}{P. McDaniel}, \bibinfo{person}{A. Sivasubramaniam}, {and}
  \bibinfo{person}{G. Tan}.} \bibinfo{year}{2020}\natexlab{}.
\newblock \showarticletitle{{IoTRepair}: Systematically Addressing Device
  Faults in Commodity IoT}. In \bibinfo{booktitle}{\emph{IoTDI}}.
\newblock


\bibitem[OLSR.org(2021)]%
        {olsrd}
\bibfield{author}{\bibinfo{person}{OLSR.org}.} \bibinfo{year}{2021}\natexlab{}.
\newblock \bibinfo{title}{Optimized Link State Routing Protocol}.
\newblock \bibinfo{howpublished}{\url{https://tinyurl.com/olsrd-wiki}}.
\newblock


\bibitem[Perkins et~al\mbox{.}(2003)]%
        {aodv}
\bibfield{author}{\bibinfo{person}{C. Perkins}, \bibinfo{person}{E.
  Belding-Royer}, {and} \bibinfo{person}{S. Das}.}
  \bibinfo{year}{2003}\natexlab{}.
\newblock \bibinfo{title}{RFC3561: Ad hoc on-demand distance vector (AODV)
  routing}.
\newblock
\newblock


\bibitem[Poirot et~al\mbox{.}(2019)]%
        {wirelesspaxos}
\bibfield{author}{\bibinfo{person}{V. Poirot}, \bibinfo{person}{B. Al~Nahas},
  {and} \bibinfo{person}{O. Landsiedel}.} \bibinfo{year}{2019}\natexlab{}.
\newblock \showarticletitle{Paxos Made Wireless: Consensus in the Air.}. In
  \bibinfo{booktitle}{\emph{EWSN}}.
\newblock


\bibitem[Research(2022)]%
        {iotagrimarket}
\bibfield{author}{\bibinfo{person}{Emergen Research}.}
  \bibinfo{year}{2022}\natexlab{}.
\newblock \bibinfo{title}{Internet of Things In Agriculture Market, By System
  (Automation and Control Systems, Sensing and Monitoring Devices, Livestock
  Monitoring Hardware, Fish Farming Hardware), By Application, and By Region
  Forecast to 2030}.
\newblock
\newblock


\bibitem[Retkowitz and Kulle(2009)]%
        {DepManagement}
\bibfield{author}{\bibinfo{person}{D. Retkowitz} {and} \bibinfo{person}{S.
  Kulle}.} \bibinfo{year}{2009}\natexlab{}.
\newblock \showarticletitle{Dependency management in smart homes}. In
  \bibinfo{booktitle}{\emph{DAIS}}.
\newblock


\bibitem[Ryu et~al\mbox{.}(2015)]%
        {ryu2015design}
\bibfield{author}{\bibinfo{person}{M. Ryu}, \bibinfo{person}{J. Yun},
  \bibinfo{person}{T. Miao}, \bibinfo{person}{I.-Y. Ahn},
  \bibinfo{person}{S.-C. Choi}, {and} \bibinfo{person}{J. Kim}.}
  \bibinfo{year}{2015}\natexlab{}.
\newblock \showarticletitle{Design and implementation of a connected farm for
  smart farming system}. In \bibinfo{booktitle}{\emph{Sensors}}. IEEE.
\newblock


\bibitem[Samsung Smart Things(2012)]%
        {SmartThings}
Samsung Smart Things \bibinfo{year}{2012}\natexlab{}.
\newblock \bibinfo{title}{Samsung {S}mart{T}hings}.
\newblock \bibinfo{howpublished}{\url{https://www.smartthings.com/}}.
\newblock


\bibitem[Shahid et~al\mbox{.}(2021)]%
        {shahid2021}
\bibfield{author}{\bibinfo{person}{H. Shahid}, \bibinfo{person}{M.~A. Shah},
  \bibinfo{person}{A. Almogren}, \bibinfo{person}{H.~A. Khattak},
  \bibinfo{person}{I.~U. Din}, {and} \bibinfo{person}{C. Kumar, N.and~Maple}.}
  \bibinfo{year}{2021}\natexlab{}.
\newblock \showarticletitle{Machine Learning-Based Mist Computing Enabled
  Internet of Battlefield Things}.
\newblock \bibinfo{journal}{\emph{ACM TOIT}} \bibinfo{volume}{21},
  \bibinfo{number}{4} (\bibinfo{year}{2021}).
\newblock


\bibitem[Shen et~al\mbox{.}(2016)]%
        {Beam}
\bibfield{author}{\bibinfo{person}{C. Shen}, \bibinfo{person}{R.~P. Singh},
  \bibinfo{person}{A. Phanishayee}, \bibinfo{person}{A. Kansal}, {and}
  \bibinfo{person}{R. Mahajan}.} \bibinfo{year}{2016}\natexlab{}.
\newblock \showarticletitle{Beam: Ending monolithic applications for connected
  devices}. In \bibinfo{booktitle}{\emph{ATC}}. \bibinfo{pages}{143--157}.
\newblock


\bibitem[Sivakumar et~al\mbox{.}(2021)]%
        {SMG21}
\bibfield{author}{\bibinfo{person}{Arun~Narenthiran Sivakumar},
  \bibinfo{person}{Sahil Modi}, \bibinfo{person}{Mateus~Valverde Gasparino},
  \bibinfo{person}{Che Ellis}, \bibinfo{person}{Andres Eduardo~Baquero
  Velasquez}, \bibinfo{person}{Girish Chowdhary}, {and}
  \bibinfo{person}{Saurabh Gupta}.} \bibinfo{year}{2021}\natexlab{}.
\newblock \showarticletitle{Learned Visual Navigation for Under-Canopy
  Agricultural Robots}. In \bibinfo{booktitle}{\emph{Robotics: Science and
  Systems}}.
\newblock


\bibitem[Song et~al\mbox{.}(2008)]%
        {song2008wirelesshart}
\bibfield{author}{\bibinfo{person}{J. Song}, \bibinfo{person}{S. Han},
  \bibinfo{person}{A. Mok}, \bibinfo{person}{D. Chen}, \bibinfo{person}{M.
  Lucas}, \bibinfo{person}{M. Nixon}, {and} \bibinfo{person}{W. Pratt}.}
  \bibinfo{year}{2008}\natexlab{}.
\newblock \showarticletitle{WirelessHART: Applying wireless technology in
  real-time industrial process control}. In \bibinfo{booktitle}{\emph{RTAS}}.
\newblock


\bibitem[Stoica et~al\mbox{.}(2001)]%
        {chord}
\bibfield{author}{\bibinfo{person}{I. Stoica}, \bibinfo{person}{R. Morris},
  \bibinfo{person}{D. Karger}, \bibinfo{person}{M.~F. Kaashoek}, {and}
  \bibinfo{person}{H. Balakrishnan}.} \bibinfo{year}{2001}\natexlab{}.
\newblock \showarticletitle{Chord: A scalable peer-to-peer lookup service for
  internet applications}.
\newblock \bibinfo{journal}{\emph{SIGCOMM}} \bibinfo{volume}{31},
  \bibinfo{number}{4} (\bibinfo{year}{2001}).
\newblock


\bibitem[Su et~al\mbox{.}(2014)]%
        {Wukong}
\bibfield{author}{\bibinfo{person}{P.~H Su}, \bibinfo{person}{C.-S. Shih},
  \bibinfo{person}{J.~Y.-J. Hsu}, \bibinfo{person}{K.-J. Lin}, {and}
  \bibinfo{person}{Y.-C. Wang}.} \bibinfo{year}{2014}\natexlab{}.
\newblock \showarticletitle{Decentralized fault tolerance mechanism for
  intelligent IoT/M2M middleware}. In \bibinfo{booktitle}{\emph{WF-IoT}}.
\newblock


\bibitem[Turchet et~al\mbox{.}(2018)]%
        {turchet2018internet}
\bibfield{author}{\bibinfo{person}{L. Turchet}, \bibinfo{person}{C. Fischione},
  \bibinfo{person}{G. Essl}, \bibinfo{person}{D. Keller}, {and}
  \bibinfo{person}{M. Barthet}.} \bibinfo{year}{2018}\natexlab{}.
\newblock \showarticletitle{Internet of musical things: Vision and challenges}.
\newblock \bibinfo{journal}{\emph{IEEE access}}  \bibinfo{volume}{6}
  (\bibinfo{year}{2018}).
\newblock


\bibitem[Vasisht et~al\mbox{.}(2017)]%
        {farmbeats}
\bibfield{author}{\bibinfo{person}{D. Vasisht}, \bibinfo{person}{Z.
  Kapetanovic}, \bibinfo{person}{J. Won}, \bibinfo{person}{X. Jin},
  \bibinfo{person}{R. Chandra}, \bibinfo{person}{S. Sinha}, \bibinfo{person}{A.
  Kapoor}, \bibinfo{person}{M. Sudarshan}, {and} \bibinfo{person}{S.
  Stratman}.} \bibinfo{year}{2017}\natexlab{}.
\newblock \showarticletitle{{FarmBeats}: An {IoT} Platform for {Data-Driven}
  Agriculture}. In \bibinfo{booktitle}{\emph{NSDI}}.
\newblock


\bibitem[Wang et~al\mbox{.}(2022)]%
        {wang2022iot}
\bibfield{author}{\bibinfo{person}{L. Wang}, \bibinfo{person}{A. Hamad}, {and}
  \bibinfo{person}{V. Sakthivel}.} \bibinfo{year}{2022}\natexlab{}.
\newblock \showarticletitle{IoT assisted machine learning model for warehouse
  management}.
\newblock \bibinfo{journal}{\emph{Journal of Interconnection Networks}}
  \bibinfo{volume}{22}, \bibinfo{number}{Supp02} (\bibinfo{year}{2022}).
\newblock


\bibitem[Xu et~al\mbox{.}(2021)]%
        {wchain}
\bibfield{author}{\bibinfo{person}{Mi. Xu}, \bibinfo{person}{C. Liu},
  \bibinfo{person}{Y. Zou}, \bibinfo{person}{F. Zhao}, \bibinfo{person}{J. Yu},
  {and} \bibinfo{person}{X. Cheng}.} \bibinfo{year}{2021}\natexlab{}.
\newblock \showarticletitle{{wChain}: A Fast Fault-Tolerant Blockchain Protocol
  for Multihop Wireless Networks}.
\newblock \bibinfo{journal}{\emph{TWC}} (\bibinfo{year}{2021}).
\newblock


\bibitem[Yang et~al\mbox{.}(2022)]%
        {medley}
\bibfield{author}{\bibinfo{person}{R. Yang}, \bibinfo{person}{J. Wang},
  \bibinfo{person}{J. Hu}, \bibinfo{person}{S. Zhu}, \bibinfo{person}{Y. Li},
  {and} \bibinfo{person}{I. Gupta}.} \bibinfo{year}{2022}\natexlab{}.
\newblock \showarticletitle{Medley: A Membership Service for {IoT} Networks}.
\newblock \bibinfo{journal}{\emph{IEEE TNSM}} \bibinfo{volume}{19},
  \bibinfo{number}{3} (\bibinfo{year}{2022}).
\newblock


\bibitem[Yang et~al\mbox{.}(2005)]%
        {YangWangKravetsWiMesh05}
\bibfield{author}{\bibinfo{person}{Y. Yang}, \bibinfo{person}{J. Wang}, {and}
  \bibinfo{person}{{R. H.} Kravets}.} \bibinfo{year}{2005}\natexlab{}.
\newblock \showarticletitle{Designing routing metrics for mesh networks}.
\newblock


\bibitem[Zolotarev(1986)]%
        {zolotarev1986onedimensional}
\bibfield{author}{\bibinfo{person}{V.~M. Zolotarev}.}
  \bibinfo{year}{1986}\natexlab{}.
\newblock \bibinfo{booktitle}{\emph{One-dimensional stable distributions}}.
  \bibinfo{series}{Translations of Mathematical Monographs},
  Vol.~\bibinfo{volume}{65}.
\newblock 277 pages.
\newblock


\end{thebibliography}

\clearpage
\newpage

\begin{appendices}
\section{Appendix: Formal Analysis}
\label{app:section:analysis}

We formally analyze \sysname{}'s properties.
For the reader who wishes to skip this section, we  summarize  our findings:

\squishlist
\item {\it Inheritance}: When a k-group changes, the state held by the old leader and new leader are identical.
\item {\it Safety}: No two routines that touch an overlapping set of devices, are allowed to execute simultaneously. 
\item {\it Liveness}: No routines deadlock.
\item {\it Progress: Every routine makes progress.}
\item We also calculate the (probabilistic) availability of the system when more than $f$ devices fail. 
\squishend

\subsection{Inheritance}

\begin{customdef}{1}
Inheritance means that given a \kgroup{}, the old and new versions of the \kgroup{}---before and after epoch change, or after leader failure---maintain identical state. 
\end{customdef}

For a device \kgroup{}, the state that must be transferred to the new leader includes:
the device's current availability and the lock queue (if applicable).
For a routine \kgroup{}, the new leader should receive its state which includes:
the routine stage, the IDs of devices that have already been locked (if applicable at
the time), and the IDs of devices that have already been released (if applicable at the time).

\begin{customlemma}{1}
\label{app:lemma:inheritance}
Inheritance is guaranteed after the state transfer stage when at most one node fails.
That is, the state held by an old \kgroup{} leader is the same as the state held by
the new \kgroup{} leader after the state transfer stage, under failure of at most one node.
\end{customlemma}

\begin{proof}
\label{lemma:inheritance:proof}
If there are new requests coming during the state transfer stage,
request processing will be delayed until the state transfer stage finishes.
The state transfer stage happens in two cases: leader failure and/or epoch change.

\phantomsection
\label{lemma:inheritance:case1}
\textbf{Case 1}:
When there is no epoch change, upon \kgroup{} leader $L_o$’s failure,
the new leader $L_n$ sends a message to all \kgroup{} members requesting their local \kgroup{} state.
All the members reply with their local state back to $L_n$.
$L_n$ includes a state entry $e$ in the recreated state
if and only if there is at least 1 received state that contains $e$.
More specifically for the device \kgroups{}, which may include a lock queue,
every routine request in the lock queue comes with a sequence number
denoting when it reached $D_k$’s \kgroup{} leader.
$L_n$ includes a routine entry $e$ in the recreated lock queue
if and only if there is at least one received queue that contain $e$.
Since each routine entry $e$ in the lock queue was replicated
by at least a quorum of \kgroup{} members before the old leader $L_o$’s failure,
$L_n$ is guaranteed to receive the exact same set of routine requests.
$L_n$ will sort the queue based on the gathered requests’ sequence numbers.
Thus, the recreated lock queue is exactly the same as $L_o$’s latest lock queue;
which also guarantees that $L_n$'s recreated state is identical to $L_o$’s latest state.

\phantomsection
\label{lemma:inheritance:case2}
\textbf{Case 2}: Upon epoch change, the system enters state transfer stage:
the new \kgroup{}’s leader $L_n$ requests the old \kgroup{}’s leader $L_o$'s state,
to which $L_o$ replies accordingly.
There are 4 possibilities during the state transfer stage:
a) the old leader fails, b) the new leader fails, c) a non-leader member of the old \kgroup{} fails, d) a non-leader member of the new \kgroup{} fails.

\phantomsection
\label{lemma:inheritance:case2a}
\textbf{Case 2a}: If  old leader $L_o$ fails after  state transfer starts
but before state transfer completes,
the new leader $L_n$ sends a request for local states to all of the old \kgroup{}’s members and
recreates $L_o$’s lock queue as in Case~\hyperref[lemma:inheritance:case1]{1}.

\phantomsection
\label{lemma:inheritance:case2b}
\textbf{Case 2b}: If  new leader $L_n$ fails before it receives the state from  old leader $L_o$,
another new leader $L_n'$ is elected and sends a new state request to $L_o$.
$L_o$ will reply again with the \kgroup{} state.

\phantomsection
\label{lemma:inheritance:case2c}
\textbf{Case 2c}: If a non-leader member of the old \kgroup{} fails,
the state transfer will not be affected and
the new \kgroup{} leader $L_n$ will receive the correct state. 

\phantomsection
\label{lemma:inheritance:case2d}
\textbf{Case 2d}: If a non-leader member of the new \kgroup{} fails,
another member will be recruited to take its place.
The new \kgroup{} leader will distribute the state to the new member.
\end{proof}

\begin{customlemma}{2}
\label{app:lemma:inheritance:extension}
Inheritance can be guaranteed after the state transfer stage when at most $\f{}$ nodes fail.
\end{customlemma}

\begin{proof}
\label{lemma:inheritance:extension:proof}
If there are no epoch changes (Lemma~\ref{app:lemma:inheritance}~Case~\hyperref[lemma:inheritance:case1]{1}) 
or failures happen only among non-leader members during the epoch change
(Lemma~\ref{app:lemma:inheritance}~Cases~\hyperref[lemma:inheritance:case2c]{2c},~\hyperref[lemma:inheritance:case2d]{2d}),
each routine entry $e$ is maintained at at least $\f + 1 - \f = 1$ alive member.
Thus, the state can be recreated correctly as in Lemma~\ref{app:lemma:inheritance}~Case~\hyperref[lemma:inheritance:case1]{1}.

Upon epoch change, if the old leader $L_o$ fails before the new leader $L_n$ gets the state,
$L_n$ (Lemma~\ref{app:lemma:inheritance}~Case~\hyperref[lemma:inheritance:case2a]{2a}) can reconstruct the same state as $L_o$
since each state entry $e$ is maintained at at least $\f + 1 - \f = 1$ alive old \kgroup{} member.
If $L_n$ keeps failing before it gets or re-constructs the state,
the latest new leader can re-construct it similarly.

If $L_o$ is alive and $L_n$ fails during epoch change, the working process is the same as Lemma~\ref{app:lemma:inheritance}~Case~\hyperref[lemma:inheritance:case2b]{2b}.
\end{proof}

\subsection{\Safety{}}
\label{app:subsec:safety}

\begin{customthm}{1}
\label{app:theorem:safety} {\bf [Safety]} 
No two routines that conflict in devices (i.e., their touched device sets are not disjoint) can execute simultaneously, under at most one node failure. 
\end{customthm}

\begin{proof}
\label{theorem:safety:proof}
Assume routines $R_1$ and $R_2$ share devices. We analyze SLA and OLA separately. 

\phantomsection
\label{theorem:safety:sla:proof}
\noindent {\bf Sequential Lock-Acquiring Strategy (SLA): } First, we will analyze  Sequential Lock-Acquiring Strategy (Section~\ref{subsubsec:sla}).
Let $D_k$ be that shared device that has the minimum $ID$, with its \kgroup{} leader denoted as $L_k$.
If $L_k$ does not fail, by definition at most one routine can reach quorum
and therefore acquire the lock on the device.

No new requests are handled during leader transition.
Such requests are put into new leader’s ($L_n$) wait queue.
The request/quorum to add such requests to the lock queue
will be initiated only after the leader transition completes. 
Therefore the rest of the proof handles only outstanding requests that have been approved.

If the leader $L_k$ fails, several cases arise:

\phantomsection
\label{theorem:safety:sla:case1}
\textbf{Case 1}: If $L_k$ fails before it has replicated the request
from $R_1$ to a quorum of \kgroup{} nodes, there is no further operation.
$R_1$’s \kgroup{} leader resends $R_1$’s request after a timeout T of not receiving an acknowledgement.

\phantomsection
\label{theorem:safety:sla:case2}
\textbf{Case 2}: If $L_k$ fails after it has replicated $R_1$’s request to a quorum of \kgroup{} members,
the new leader $L_n$ adds $R_1$ to the lock queue {during the reconstruction phase when it queries (a quorum of) k-group members for their pieces of the state.}

\phantomsection
\label{theorem:safety:sla:case3}
\textbf{Case 3}: If $L_k$ fails after $R_1$’s request reaches the top of the lock queue,
two sub-cases occur: a) If the request has been approved by a quorum of nodes
(i.e., been forwarded to nodes, and a quorum has approved $R_1$ to acquire the lock),
the new leader $L_n$ marks $R_1$ as the approved routine and notifies $R_1$’s \kgroup{} leader.
b) If the request has not yet reached quorum (to approve the routine for lock acquisition):
then the new leader $L_n$ adds $R_1$ to the lock queue, and continues its normal processing.

\phantomsection
\label{theorem:safety:sla:case4}
\textbf{Case 4}: If $L_k$ fails after it has sent a Locked message to $R_1$’s k-group,
the new leader $L_n$: 1) re-constitutes the lock queue from the surviving nodes,
and 2) (re)notifies $R_1$’s \kgroup{} leader of its approval.

{

\phantomsection
\label{theorem:safety:ola:proof}
\noindent {\bf Optimistic Lock-Acquiring Strategy (OLA):}
For each device $D_k$ that is touched by both $R_1$ and $R_2$,
only one of the two routines' lock requests will be processed first by $D_k$'s \kgroup{}
leader $L_k$ and reach quorum for the device pre-lock, as long as $L_k$ does not fail.
By definition  only this routine will go on to lock the device if all other desired
devices are also available and each desired device's $D_k$'s $L_k$ does not fail.

If a leader $L_k$ does fail, two cases arise:

\phantomsection
\label{theorem:safety:ola:case1}
\textbf{Case 1}: If $L_k$ fails before it gets a quorum of acknowledgement for pre-lock replication,
the new leader $L_n$ will have no record of it.
$R_1$'s \kgroup{} leader will resend the request after timeout.

\phantomsection
\label{theorem:safety:ola:case2}
\textbf{Case 2}: If $L_k$ fails after it has replicated $R_1$'s pre-lock request, 
but before it hears back from $R_1$'s leader for further action (of either locking for execution or releasing due to failed pre-lock),
the new leader $L_n$ will know $R_1$ has pre-locked $D_k$ from its group member and resend pre-lock information back to $R_1$.

\phantomsection
\label{theorem:safety:ola:case3}
\textbf{Case 3}: If $L_k$ fails after it hears from $R_1$ for lock acquisition or releasing, but before it gets quorum acknowledgement for replication, $R_1$'s \kgroup{} leader will resend the request after timeout and the locking process keeps up.

\phantomsection
\label{theorem:safety:ola:case4}
\textbf{Case 4}: If $L_k$ fails after it has locked $D_k$ and notifies $R_1$'s \kgroup{}
the new leader $L_n$ 1) re-constitutes the state from the surviving nodes,
and 2) (re)notifies $R_1$’s \kgroup{} leader of its approval.
Therefore, the OLA strategy is safe as well.
}
Thus \sysname{} is safe under at most one node failure.
\end{proof}

\subsection{\Liveness{}}
\label{app:subsec:liveness}

\begin{customthm}{2}
\label{app:theorem:deadlock} {\bf [No Deadlocks]} 
No two routines are stuck in a deadlock.
\end{customthm}

\begin{proof}
\label{theorem:deadlock:proof:sla}
Each SLA and OLA violate separate necessary conditions for deadlocks. 
For SLA, routines only wait for higher-$ID$ devices than they have already locked.
This means no routine waits for lower-$ID$ devices than it has locked.
Thus, there is no ``wait for'' cycle among routines.
\phantomsection
\label{theorem:deadlock:proof:ola}
{
As for OLA, routines do not wait for any already-locked devices.
If a device is already (pre-)locked, the new routine will abort and try to acquire
its touched devices' locks at a later time. Without ``hold and wait'' there is no deadlock.
}

Leader change and failure do not affect device locks held by \sysname.
Meanwhile Lemmas~\ref{app:lemma:inheritance},~\ref{app:lemma:inheritance:extension} guarantee inheritance.
Thus  lock assignment is not influenced by leader changes and failures, i.e., there is no deadlock.
\end{proof}

Finally, we prove that \sysname{} always  makes progress. We prove the induction step via the following theorem (this theorem can be applied iteratively).

\begin{customthm}{3}
\label{app:theorem:progress} {\bf [Progress/Liveness]} 
Consider a set of executing routines $\mathcal{R}$ that are each
 still waiting to acquire 1 or more locks, and a set of devices $D$ they are waiting for. If no further routines arrive into  $\mathcal{R}$, and SLA is used for locking,  then: at least one routine from the set $\mathcal{R}$ will make progress (i.e., get access to one more devices that it desires).
\end{customthm}
\begin{proof}
Consider the highest-$ID$ device $D$ in set $D$, and the routine $R$ that currently holds the lock on device $D$. Then  $R$ cannot be waiting for any further locks (otherwise $D$ would not be the highest-$ID$ in $D$), and thus $R$ will eventually complete. At that point, the routine at the head of $D$'s queue will be granted access to $D$ and thus  will make progress. \end{proof}

{For OLA, an adversary may always be able to prevent two competing routines from making progress. However, in practice, we observed that routines make fast progress in OLA. Nevertheless, because of this, we use SLA as the default in the \sysname{} implementation.}

\subsection{System Availability}
While \sysname{} can tolerate $f$ failures, what happens if there are more than $f$ failures? With $F (>f)$ simultaneous failures, we calculate the {\it availability} $P(\F{})$ =  
the probability that \sysname{}  still works correctly, i.e., all \kgroups{} contains at least $(f+1)$ non-faulty nodes.
Assume we have $S$ \smart{}s, $G$ \kgroups{}, and each group has $k$ members.

\begin{enumerate}
    \item When $\F{} \le \f{}$, $P(F) = 100\%$
    \item When $\F{} \ge \f{}$, the {\it availability} is: 
        \begin{align*}
          P(F) &= P(\text{at most } f \text{ failures across all } k \text{-groups}) \\
               &= P(\text{at most } f \text{ failures in one } k \text{-group})^G  \\
               &= \Bigg(\sum_{i=\max(0, k+F-S)}^{f}\frac{\binom{S-F}{k-i}\times \binom{F}{i}}{\binom{S}{k}}\Bigg)^G
        \end{align*} 
\end{enumerate}

{The equation for one \kgroup{} is explained as follows. Given $F$ failures, the numerator is the number of ways to select a \kgroup{} containing  $(k-i)$ non-faulty nodes (left term), and $i$ faulty nodes (right term), while the denominator  is the number of all possible selections for a \kgroup{}. $i$ varies from $f$ down to $0$ (or a higher number $(k+F-S)$ if there are insufficient alive nodes). }

{From \Figure~\ref{subfig:prob-on-f} we observe the availability of \sysname{} drops when the number of simultaneous failures $\F{}$ becomes  $> \f{}$. However, \sysname{}  still provides 50\% of availability  when simultaneous failures $\F{}$ grow up to 9$\times f$. \Figure~\ref{subfig:prob-varyg} shows that to reach a specific percentile of availability $p$, \sysname{}'s tolerated $F$ drops very slowly as the number of groups increases---this indicates scalability with the number of groups.}

\end{appendices}

\vspace{12pt}

\end{document}